\documentclass[pra,aps]{revtex4}

\usepackage{amssymb}

\usepackage{graphicx}

\begin{document}
\title{Sample dependence of the Casimir force}
\author{V. B. Svetovoy}
\email[]{V.B.Svetovoy@el.utwente.nl}
\thanks{On leave from Yaroslavl University, Russia}
\affiliation{Transducers Science and Technology Group,
MESA+ Research Institute,
University of Twente, P.O. 217, 7500 AE Enschede, The Netherlands}
\date{\today}

\pacs{12.20.Ds, 12.20.Fv, 42.50.Lc, 73.61.At, 77.22.Ch}

\begin{abstract}
Difference between bulk material and deposited film is shown to have
an appreciable influence on the Casimir force. Analysis of the optical
data on gold films unambiguously demonstrates the sample dependence:
the dielectric functions of the films deposited in different conditions
are different on the level that cannot be ignored in high precision
prediction of the force. It is argued that the precise values of
the Drude parameters are crucial for accurate evaluation of the force.
The dielectric
function of perfect single crystalline gold is discussed. It is used to
establish the upper limit on the absolute value of the force. It is
demonstrated that the force between films is smaller than that
between bulk samples mainly due to the presence of voids and electron
scattering on the grain boundaries in the films. The minimal reduction
of the force is estimated as 2\% for the smallest separations
investigated in the most precise experiments. The other sample effects
can reduce the force further but the correction is expected to be
smaller than 1\%.
\end{abstract}

\maketitle

\section{Introduction\label{Sec1}}

The Casimir force \cite{Cas48} between uncharged metallic plates
attracts considerable attention as a macroscopic manifestation of
the quantum vacuum \cite{Mil94,Mos97,Mil01,Kar99,Bor01}. With the
development of microtechnologies that routinely allow control of the
separation between bodies smaller than 1 $\mu m$ the force became
a subject of systematic experimental investigation. Modern high
precision experiments were made using different techniques such as
torsion pendulum \cite{Lam97}, atomic force microscope (AFM) \cite
{Moh98,Har00}, microelectromechanical systems (MEMS) \cite
{Cha01,Dec03a,Dec03b} and different geometrical configurations:
sphere-plate \cite{Lam97,Har00,Dec03b}, plate-plate \cite{Bres02},
crossed cylinders \cite {Ede00}. In most cases the bodies were
covered with gold evaporated or sputter deposited to the thickness
of 100-200 nm. In the only plate-plate configuration experiment
\cite{Bres02} the bodies were covered with chromium. One of the
bodies was covered with copper in the MEMS experiment \cite
{Dec03a,Dec03b}. The root mean square (rms) roughness of deposited
metal films was varied from 30-40 nm as in the MEMS experiments
\cite {Cha01,Dec03a,Dec03b} to nearly atomically flat surfaces as
in the crossed cylinders experiment \cite{Ede00}. Relatively low
precision, 15\%, in the force measurement was reached for the
plate-plate configuration \cite{Bres02} because of the parallelism
problem. In the torsion pendulum experiment \cite {Lam97} the
force was measured with the accuracy of 5\%. In the experiments
\cite{Har00,Ede00,Cha01} it was found with 1\% precision. In the
most precise up to date experiment \cite{Dec03a,Dec03b} the
improvement was achieved due to the use of the dynamical method
\cite{Cha01,Bres02}. Additionally the change in the resonance
frequency of the mechanical oscillator was measured using the phase
jump instead of the resonance behavior of the amplitude. In this way the
force between gold and copper covered bodies was found with a relative
accuracy of 0.25\%.

To draw any conclusion from the experiments one has to predict the force
theoretically with the precision comparable with the experimental errors.
It is a real challenge to the theory because the force is material
dependent; it is very difficult to fix the material properties since
different uncontrolled factors are involved. In its original form, the
Casimir force \cite{Cas48}

\begin{equation}
F_{c}\left( a\right) =-\frac{\pi ^{2}}{240}\frac{\hbar c}{a^{4}}
\label{Fc}
\end{equation}

\noindent was calculated between the ideal metals. It depends only on the
fundamental constants and the distance between the plates $a$. The force
between real materials was found for the first time by Lifshitz \cite
{Lif56,LP9}. The material properties enter in the Lifshitz formula via the
dielectric function $\varepsilon \left( i\zeta \right) $ at imaginary
frequencies $\omega=i\zeta $. Correction to the expression (\ref{Fc}) is
significant at the separations between bodies smaller than 1 $\mu m$.
To calculate the force the Lifshitz formula is used with the
optical data taken from the handbooks \cite{HB1,HB2}. The data are available
only up to some low-frequency cutoff $\omega _{c}$. For good metals such as
$Au,$ $Al,$ $Cu$ the data can be extrapolated to the lower frequencies
$\omega<\omega_c$ with the Drude dielectric function

\begin{equation}
\varepsilon \left( \omega \right) =1-\frac{\omega _{p}^{2}}{\omega \left(
\omega +i\omega _{\tau }\right) },  \label{Drude}
\end{equation}

\noindent which includes two parameters: the plasma frequency
$\omega _{p}$ and the relaxation frequency $\omega _{\tau }$.
These parameters can be extracted from the optical data at the
lowest accessible frequencies. The exact values of the Drude
parameters are very important for the precise evaluation of the force.

When two plates are separated by the
distance $a$, one can introduce a characteristic imaginary frequency
$\zeta_{ch}=c/2a$ of electromagnetic field fluctuations in the gap. The
important fluctuations are those that have frequency $\zeta \sim
\zeta _{ch}$. For $a\sim 100\ nm$ the characteristic frequency is
in the near infrared region. This qualitative consideration is often
continued to the real frequency domain \cite{Moh98,Kli99,Kli00,Che04}
with the real characteristic frequency $\omega_{ch}=c/2a$. Because
$\omega_{ch}$ is in the near infrared it is stated that the
low-frequency behavior of the dielectric function is not significant
for the Casimir force. It is not difficult to see that this consideration
is wrong \cite{Sve00b,Ian03,Ian04}. The force depends on the dielectric
function at imaginary frequencies $\varepsilon \left( i\zeta \right) $.
With the help of the dispersion relation $\varepsilon \left(
i\zeta \right) $ can be expressed via the observable function
$\varepsilon ^{\prime \prime }\left( \omega \right)=
Im \varepsilon \left( \omega \right) $:

\begin{equation}
\varepsilon \left( i\zeta \right) -1=\frac{2}{\pi }\int\limits_{0}^{\infty }%
d\omega\frac{\omega \varepsilon ^{\prime \prime }\left( \omega \right) }{\omega
^{2}+\zeta ^{2}}.  \label{K-K}
\end{equation}

\noindent Because for metals $\varepsilon ^{\prime \prime }\left( \omega \right)$
is large at low frequencies, the main contribution in the
integral in Eq. (\ref{K-K}) is given by the low frequencies even if $\zeta $ is
in the visible range. The characteristic
frequency is well defined on the imaginary axis $\zeta
_{ch}=c/2a$ but it is wrong to introduce it in the real frequency
domain. This aspect is carefully discussed in this paper.

Two different procedures to get the Drude parameters were
discussed in the literature. In the first approach \cite{Lam00,Kli00}
the plasma frequency was fixed independently on the optical data
assuming that each atom gives one conduction electron (for $Au$) with
the effective mass equal to the mass of free electron. The optical
data at the lowest frequencies were used then to find $\omega
_{\tau }$ with the help of Eq. (\ref{Drude}). In this way the
parameters $\omega _{p}=9.0\ eV$ and $\omega _{\tau }=0.035\ eV$
have been found. In the second approach \cite{Bos00b,Sve00a,Sve00b} no
assumptions were made and both of the parameters were extracted
from the low-frequency optical data by fitting them with
Eq. (\ref{Drude}). When the best data from Ref. \cite{HB2} were used the
result \cite{Sve00b} was close to that found by the first approach, but
using different sources for the optical data collected in Ref.
\cite{HB2} an appreciable difference was found \cite{Sve00a,Sve00b}.
This difference was attributed to the defects in the metallic
films which appear as the result of the deposition process. It was
indicated that the density of the deposited films is typically
smaller and the resistivity is larger than the corresponding
values for the bulk material. However, it was difficult to make
definite conclusions without analysis of significant amount of
the data on the deposited metallic films. In this paper this
analysis is performed and it is proven that there is a genuine
sample dependence of optical properties of gold. First of all it
means that the gold films prepared in different ways will have
different Drude parameters. The change in the optical properties
will lead to the sample dependence of the Casimir force, which is
carefully investigated in this paper. The main ideas were outlined
before \cite{Sve03b} with the conclusion that the absolute value
of the Casimir force between gold films is of about 2\% smaller
than between bulk material at the smallest separations
investigated in the experiments \cite{Har00,Dec03b}. Here we come
to the same conclusion but present much more details and facts.

The paper is organized as follows. In Sec. \ref{Sec2} it is
explained qualitatively and quantitatively why the exact values of
the Drude parameters are crucial for the precise calculation of
the Casimir force. Existing optical data for gold are reviewed and
analyzed in Sec. \ref{Sec3}. It is unambiguously demonstrated that
there is the sample dependence of the optical properties of gold films.
In Sec. \ref {Sec4} the dielectric function of the perfect gold
single crystal is discussed and the upper limit on the Casimir
force is deduced. In Sec. \ref {Sec5} a simple two parameter
model is proposed allowing us to take into account the main reasons
for the sample dependence of the Casimir force: voids and grains
in the films. The volume fraction of voids and the mean grain size are
bounded from the available optical data and the smallest sample
correction to the force is estimated. In Sec. \ref{Sec6} necessary
experimental information on the films that has to be known for
the precise evaluation of the force is briefly discussed.
Comparison between the AFM experiment and prediction of this paper
is given. Our conclusions are summarized in the last section.

\section{Importance of the precise values of the Drude
parametrs\label{Sec2}}

Let us consider the Casimir force between similar metallic plates
separated by a distance $a$ at a temperature $T$. It depends on
the material optical properties via the dielectric function
$\varepsilon \left( \omega \right) $ at imaginary frequency
$\omega =i\zeta $ \cite{LP9}:

\begin{equation}
F_{pp}\left( a\right) =-\frac{k_{B}T}{\pi }\sum_{n=0}^{\infty
}{}^{\prime
}\int\limits_{0}^{\infty }dqqk_{0}\left[ \left( r_{s}^{-2}\exp \left(
2ak_{0}\right) -1\right) ^{-1}+\left( r_{p}^{-2}\exp \left( 2ak_{0}\right)
-1\right) ^{-1}\right] ,  \label{Fpp}
\end{equation}

\noindent where ${\bf q}$ is the wave vector along the plates ($q=\left|
{\bf q}\right| $). This formula includes the reflection coefficients for two
polarization states $s$ and $p$ which are defined as

\begin{equation}
r_{s}=\frac{k_{0}-k_{1}}{k_{0}+k_{1}},\quad r_{p}=\frac{\varepsilon
\left(
i\zeta _{n}\right) k_{0}-k_{1}}{\varepsilon \left( i\zeta _{n}\right)
k_{0}+k_{1}}.  \label{rsp}
\end{equation}

\noindent Here $k_{0}$ and $k_{1}$ are the normal components of
the wave vectors in vacuum and metal, respectively:

\begin{equation}
k_{0}=\sqrt{\zeta _{n}^{2}/c^{2}+q^{2}},\quad
k_{1}=\sqrt{\varepsilon \left(
i\zeta _{n}\right) \zeta _{n}^{2}/c^{2}+q^{2}}.  \label{k01}
\end{equation}

\noindent The imaginary frequencies here are the Matsubara frequencies

\begin{equation}
\zeta _{n}=\frac{2\pi k_{B}T}{\hbar }n.  \label{Mats}
\end{equation}

For most of the experimental configurations the force has to be
calculated between sphere and plate. While the sphere radius $R$
is much larger than the minimal separation $a$, the force can be
approximately calculated with the help of the proximity force
approximation \cite{Der57} that gives the following result:

\begin{equation}
F_{sp}\left( a\right) =k_{B}TR\sum_{n=0}^{\infty }{}^{\prime
}\int\limits_{0}^{\infty }dqq\left[ \ln \left( 1-r_{s}^{2}\exp \left(
-2ak_{0}\right) \right) +\ln \left( 1-r_{p}^{2}\exp \left( -2ak_{0}\right)
\right) \right] .  \label{Fsp}
\end{equation}

\noindent This approximation is good with the precision $\sim a/R$ that is
typically better or on the level of 0.1\%.

The material dielectric function $\varepsilon (i\zeta )$ is the main source
of uncertainty for precise prediction of the Casimir force. This function
cannot be measured directly but it can be expressed via the measurable function
$\varepsilon \left( \omega \right) =\varepsilon ^{\prime }\left(
\omega \right) +i\varepsilon ^{\prime \prime }\left( \omega
\right) $ with the help of the well known dispersion relation (\ref{K-K}).
The equation (\ref{K-K}) shows that
the Casimir force is intimately connected with the dissipation in
the material given by $\varepsilon ^{\prime \prime }\left( \omega
\right) $ that is, indeed, the result of the fluctuation-dissipation
theorem (see discussion in Ref. \cite{Sve04a}).

The exponent $\exp \left( 2ak_{0}\right) $ in Eqs. (\ref{Fpp}),
(\ref{Fsp}) shows that the main contribution to the force comes
from the region $\sqrt{\zeta ^{2}/c^{2}+q^{2}}\sim 1/2a$. For this
reason one can say that the field fluctuations with frequencies
nearby the characteristic imaginary frequency $\zeta _{ch}=c/2a$
give the main contribution to the force. This frequency is in the
infrared region for typical separations $a=100-500\,nm$
investigated in the experiments. Therefore, for precise
calculation of the force we have to know the dielectric function
at imaginary frequencies $\zeta \sim \zeta _{ch}$ with the highest
possible accuracy. This qualitative analysis is often extended to
the real frequencies $\omega $ \cite{Moh98,Kli99,Kli00,Che04}
that is, of course, not true. As follows from the dispersion
relation (\ref{K-K}) real frequencies $\omega $ essentially
different from $c/2a$ dominate in $\varepsilon \left( i\zeta
_{ch}\right) $. Suppose that the material can be described by the
Drude dielectric function (\ref{Drude}). Then the integrand in
Eq. (\ref{K-K}) has the form $\omega _{p}^{2}\omega _{\tau }(\omega
^{2}+\omega _{\tau }^{2})^{-1}\left( \omega ^{2}+\zeta
_{ch}^{2}\right) ^{-1}$. This function is significant in the range
$\omega \lesssim \omega _{\tau }\ll \zeta_{ch}$. Since $\omega _{\tau }$
corresponds to the far infrared region, we have to
know the function $\varepsilon ^{\prime \prime }\left( \omega \right) $
with the highest precision in the far infrared but not in the infrared.
Let us consider this important conclusion in detail.

Throughout this paper the energy units will be used to measure the
frequency. The following transition coefficients are helpful to
transform it in the wavelength or in $rad/s$:

\begin{equation}
1\ eV=1.241\ \mu m^{-1},\quad 1\ eV=1.519\cdot
10^{15}\ \frac{rad}{s}.
\label{units}
\end{equation}

\begin{figure}[tbp]
\includegraphics[width=8.6cm]{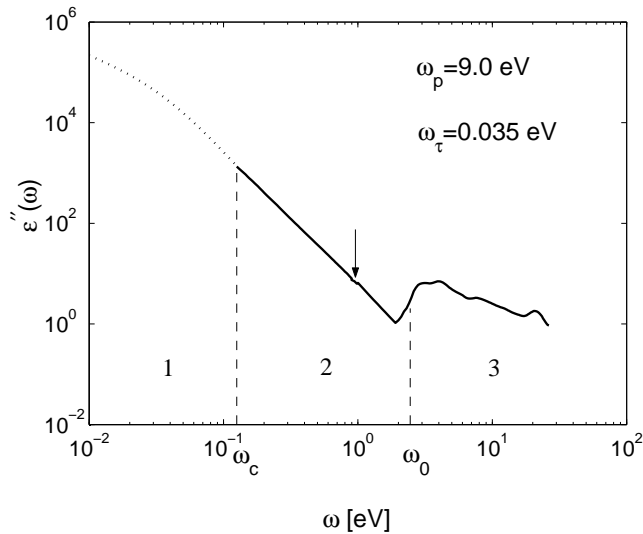}\newline
\caption{Handbook data for $Au$ \cite{HB1} (solid line). The arrow
separates the data from the original sources \cite{Dol65} (on the
left) and \cite{The70} (on the right). The dotted line shows the
extrapolation to low frequencies with the Drude parameters
indicated in the upper right corner. The cutoff frequency $\omega
_{c}$ and the edge of the interband transition $\omega _{0}$
separate three frequency domains marked as {\it 1}, {\it 2}, and
{\it 3}.} \label{fig1}
\end{figure}

\noindent For precise calculation with Eq. (\ref{Fpp}) or (\ref{Fsp}) the
dielectric function $\varepsilon ^{\prime \prime }\left( \omega
\right) $ is usually taken from the handbooks. Fig. \ref{fig1} comprises
the handbook data for gold \cite{HB1}. The solid line shows the actual data
taken from two original sources: the
points to the right of the arrow are those by Th\`{e}ye \cite{The70} and
to the left by Dold and Mecke \cite{Dol65}. For frequencies smaller than the
cutoff frequency $\omega _{c}$ ($0.125\ eV$ for these data) the data
are unavailable and $\varepsilon ^{\prime \prime }\left( \omega \right) $ has
to be extrapolated into the region $\omega <\omega _{c}$. This is a delicate
procedure because the contribution of the low frequencies dominates in $%
\varepsilon \left( i\zeta \right) $. In the field of the Casimir effect the
following procedure has been adopted \cite{Lam00}. The plasma frequency
is calculated using the relation

\begin{equation}
\omega _{p}^{2}=\frac{Ne^{2}}{\varepsilon _{0}m_{e}^{\ast }},
\label{Omp}
\end{equation}

\noindent where $N$ is the number of conduction electrons per unit
volume, $e $ is the charge and $m_{e}^{\ast }$ is the effective mass of electron.
Assuming that each atom gives one conduction electron, that the effective
mass coincides with the mass of free electron and using the bulk density of
$Au$ it was found $\omega _{p}=9.0\ eV$. With this value of the
plasma frequency the best fit of available low-frequency data with
the Drude function (\ref{Drude}) gave for the relaxation frequency
$\omega _{\tau }=0.035\ eV$ \cite{Lam00}. The dotted line in Fig.
\ref{fig1} shows the Drude extrapolation with these parameters.

One can separate three frequency regions in Fig. \ref{fig1} marked as {\it
1}, {\it 2}, and {\it 3}. In the high energy domain $\omega >\omega _{0}$,
where $\omega _{0}\approx 2.45\ eV$ \cite{The70} is the edge of interband
absorption, the Casimir force only weakly depends on the behavior of the
dielectric function. However, as we will see in the
next section, analysis of the interband absorption can give valuable
information about voids in the metallic films.
The region {\it 2} extends from the cutoff frequency $\omega _{c}$
to the absorption edge $\omega _{0}$. The curve in this region is
also sensitive to the sample quality. It gives perceptible
contribution to $\varepsilon \left( i\zeta \right) $ but what is more
important is that this domain completely defines the Drude parameters.
These parameters will be used to extrapolate the curve to the low-frequency region
{\it 1}, $\omega<\omega _{c}$. The low-frequency region gives the
main contribution to $\varepsilon \left( i\zeta \right) $ and for
this reason the precise values of the Drude parameters are very important.

Let us write the dielectric function at imaginary frequencies in the
following form:

\begin{equation}
\varepsilon \left( i\zeta \right) =1+\varepsilon _{1}\left( i\zeta \right)
+\varepsilon _{2}\left( i\zeta \right) +\varepsilon _{3}\left( i\zeta
\right) ,  \label{split}
\end{equation}

\noindent where the indexes 1,2 and 3 indicate the integration range
in Eq. (\ref{K-K}) that are $0\leq \omega <\omega _{c}$, $\omega
_{c}\leq\omega <\omega _{0}$, and $\omega _{0}\leq \omega <\infty $,
respectively. $\varepsilon _{1}$ can be calculated explicitly using
the Drude function (\ref{Drude}). It gives

\begin{equation}
\varepsilon _{1}\left( i\zeta \right) =\frac{2}{\pi }\frac{\omega
_{p}^{2}}{\zeta ^{2}-\omega _{\tau }^{2}}\left[ \tan ^{-1}\left( \frac{\omega
_{c}}{\omega _{\tau }}\right) -\frac{\omega _{\tau }}{\zeta }\tan ^{-1}\left(
\frac{\omega _{c}}{\zeta }\right) \right] .  \label{eps1}
\end{equation}

\begin{figure}[tbp]
\includegraphics[width=8.6cm]{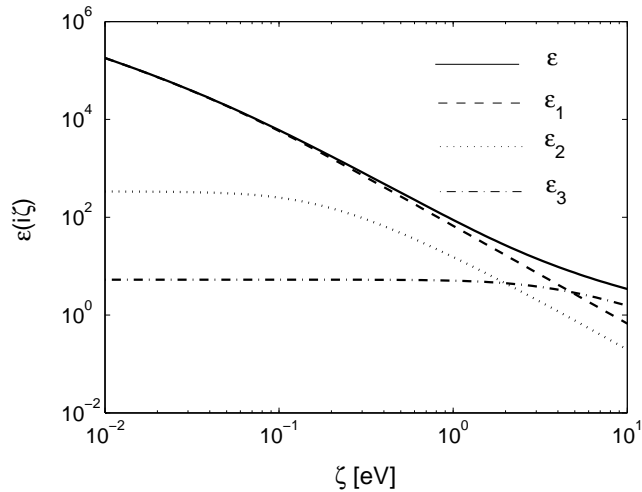}\newline
\caption{Contributions of different real frequency domains to the
dielectric function on the imaginary axis $\varepsilon(i\zeta)$.
The low-frequency ($\omega<\omega_c$) contribution
$\varepsilon_{1}(i\zeta)$ clearly dominates while $\zeta<5\ eV$.}
\label{fig2}
\end{figure}

\noindent Two other functions $\varepsilon _{2}$ and $\varepsilon _{3}$
have to be calculated numerically. The result for all functions including $%
\varepsilon \left( i\zeta \right) $ is shown in Fig. \ref{fig2}.
One can clearly see that $\varepsilon _{1}\left( i\zeta \right) $
dominates in the dielectric function at imaginary frequencies up
to $\zeta \approx 5\ eV$. It means that for all experimentally
investigated situations the region {\it 1,} where the extrapolated
optical data are used, gives the main contribution to $\varepsilon
\left( i\zeta \right) $. For example, for the separation between
bodies $a=100\ nm$ the characteristic imaginary frequency is
$\zeta _{ch}=0.988\ eV$. At this frequency the contributions of
different domains to $\varepsilon \left( i\zeta _{ch}\right) $
are $\varepsilon _{1}=68.42$, $\varepsilon _{2}=15.65$, and $\varepsilon _{3}=5.45$.

In the following section it will be demonstrated that the dielectric
function of the films used for the force measurement can deviate
considerably from the handbook data. The main reason for this is
genuine sample dependence of the optical properties. Here let us
consider the question how sensitive is the Casimir force to this
deviation. To this end consider small variations $\Delta \varepsilon _{n} $
of the dielectric function $\varepsilon _{n}\equiv \varepsilon
(i\zeta _{n})$. The relative change in the force can be written
via the relative variation of $\varepsilon _{n}$ as

\begin{equation}
\frac{\Delta F}{F}=\frac{1}{F}\sum_{n=0}^{\infty }{}^{\prime }\left(
\frac{%
\partial F_{n}}{\partial \varepsilon _{n}}\varepsilon _{n}\right) \frac{%
\Delta \varepsilon _{n}}{\varepsilon _{n}},  \label{dF}
\end{equation}

\noindent where $F_{n}$ is the $n$-th term in the sum (\ref{Fpp}) or
(\ref{Fsp}) and $F$ is the total force. To see how important is the change of the
dielectric function at real frequencies, it is more convenient to introduce
instead of $\Delta \varepsilon _{n}/\varepsilon _{n}$ the relative variations
for each frequency domain {\it 1}, {\it 2}, or {\it 3}:

\begin{equation}
\frac{\Delta \varepsilon _{n}}{\varepsilon _{n}}=\frac{\varepsilon
_{n1}}{\varepsilon _{n}}\Delta _{n1}+\frac{\varepsilon _{n2}}{\varepsilon
_{n}}\Delta _{n2}+\frac{\varepsilon _{n3}}{\varepsilon _{n}}\Delta _{n3}
\label{Delta}
\end{equation}

\noindent with

\begin{equation}
\Delta _{nj}=\frac{2}{\pi \varepsilon _{nj}}\int_{D_{j}}d\omega
\frac{\omega
\Delta \varepsilon ^{\prime \prime }\left( \omega \right) }{\omega
^{2}+\zeta _{n}^{2}},\qquad j=1,2,3.  \label{deli}
\end{equation}

\noindent Here $\varepsilon _{nj}$ is defined as $\varepsilon _{j}\left(
i\zeta _{n}\right) $, $\Delta \varepsilon ^{\prime \prime }\left( \omega
\right) $ is the measured deviation at real frequencies and $D_{j}$ is a
particular frequency interval coinciding with the domain $j$. With these
definitions the relative change in the force can be presented as a linear
combination of $\Delta _{nj}$:

\begin{equation}
\frac{\Delta F}{F}=\frac{1}{F}\sum_{n=0}^{\infty }{}^{\prime }\left[
\left(
\frac{\partial F_{n}}{\partial \varepsilon _{n}}\varepsilon _{n1}\right)
\Delta _{n1}+\left( \frac{\partial F_{n}}{\partial \varepsilon _{n}}%
\varepsilon _{n2}\right) \Delta _{n2}+\left( \frac{\partial F_{n}}{\partial
\varepsilon _{n}}\varepsilon _{n3}\right) \Delta _{n3}\right] .
\label{DF}
\end{equation}

\begin{figure}[tbp]
\includegraphics[width=8.6cm]{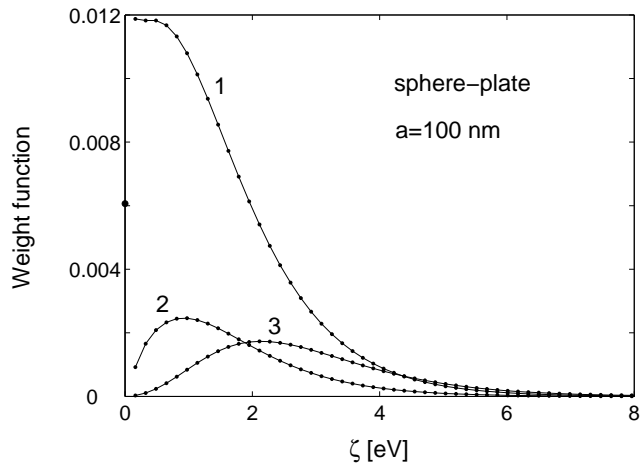}\newline
\caption{The weight functions defined by the Eq. (\ref{DF}) for
different frequency domains. The point on the ordinate axis
shows the weight of the $n=0$ term (see Eq. (\ref{zterm})), which
exists only for the plasma model prescription \cite{Bor00}.}
\label{fig3}
\end{figure}

\noindent The coefficients in front of $\Delta _{j}$ can be called the
weight functions. They describe the weight with which domain $j$
contributes
to the force. These weight functions are shown in Fig. \ref{fig3} for the
separation $a=100\ nm$. One can clearly see that the weight for the
domain {\it 1} is the largest. When $a$ increases the weight functions become
more narrow and the domains {\it 2} and {\it 3} become less
important. In the next section the relative variation of $\varepsilon
_{nj}$ for real gold films will be discussed where more detailed
information on $\Delta _{nj}$ will be presented.

The zero term $n=0$ in Eqs. (\ref{dF})-(\ref{DF}) has to be described
separately. In the Lifshitz formula (\ref{Fpp}) or (\ref{Fsp}) the $n=0$
term rose controversy in the literature \cite
{Bos00a,Sve00b,Bor00,Gen00,Sve01,Kli01,Bre02,Hoy03}. It is still an
unresolved problem in the field of the Casimir effect. The point of
disagreement between different authors is the value of the reflection
coefficient $r_{s}$ in the limit $\zeta \rightarrow 0$. Bostr\"{o}m and
Sernelius \cite{Bos00a} found $r_{s}=0$ applying the Drude behavior to
$\varepsilon \left( i\zeta \right) $ at $\zeta \rightarrow 0$. \ Svetovoy and
Lokhanin \cite{Sve00a,Sve00b,Sve01} got $r_{s}=1$ demanding
continuous transition to the ideal metal at $\zeta \rightarrow 0$. Bordag et al.
\cite{Bor00} have used the dielectric function of the plasma model in the limit
$\zeta \rightarrow 0$ and found

\begin{equation}
r_{s}=\frac{q-\sqrt{\omega _{p}^{2}/c^{2}+q^{2}}}{q+\sqrt{\omega
_{p}^{2}/c^{2}+q^{2}}}.  \label{rsplasma}
\end{equation}

\noindent There is no problem with $r_{p}$ coefficient which is always
$r_{p}=1$ in the $\zeta \rightarrow 0$ limit.

In the first two approaches $r_{s}$ and $r_{p}$ do not depend on the
material and the zero term should be omitted in Eqs. (\ref{dF})-
(\ref{DF}). In the plasma model approach \cite{Bor00} $r_{s}$ depends on the
plasma frequency and the $n=0$ term will change when $\omega _{p}$ varies.
The contribution of the zero term to Eq. (\ref{DF}) will be

\begin{equation}
\left( \frac{\Delta F}{F}\right) _{zero\ term}=\frac{1}{2F}\left( \frac{%
\partial F_{0}}{\partial \omega _{p}}\omega _{p}\right) \frac{\Delta
\omega_{p}}{\omega _{p}}.  \label{zterm}
\end{equation}

\noindent In Fig. \ref{fig3} the coefficient in front of $\Delta \omega
_{p}/\omega _{p}$ is shown by the dot on the ordinate axis.

The main conclusion that can be drawn from the analysis above is that the
deviation of the Drude parameters $\omega _{p}$ and $\omega _{\tau }$
from their handbook values gives the largest contribution to
the change of the Casimir force.

\section{Optical data for gold and analysis\label{Sec3}}

In this section we are going to show that the optical properties of
gold films depend on the details of the deposition method and
following treatment to the extent that cannot be ignored in the
calculation of the Casimir force. Optical properties of gold were
extensively investigated in 50-70th. Today most of the researches
are interested in gold nanoclusters but these data are inappropriate
for our purposes. There are only a few exceptions. Reflectivity of
evaporated gold films was investigated \cite{Sot03} in the
wavelength range $0.3-50$ $\mu m$ to analyze its dependence on the
mean grain size. Since only normal reflectance has been measured,
one cannot get the information on both real and imaginary part of
the dielectric function but these data are very helpful. The
reflectivity of gold in submillimeter range, $513\ \mu m$, was
measured with high precision \cite{Gat95}. In Ref. \cite{Wan98} the
films evaporated on glass with two different methods were
characterized ellipsometrically in the photon energy range $1.5-4.5\
eV$. This work allows one to compare the interband absorption with
the older measurements. In addition, the optical properties of films
deposited with the magnetron dc sputtering and with the filtered arc
process \cite{Ben99} were measured at the photon energies $1.5-3.5\
eV$ and compared with each other. One can indicate also two recent
studies \cite{An02,Xia00} where new ellipsometric equipment was
tested but the preparation of the films was poorly described. We
will use the recent papers where it is possible but most of the data
will be taken from the old works. In many old works the importance
of sample preparation methods was recognized and carefully
discussed. Complete bibliography of the old publications up to 1981
one can find in Ref. \cite{Wea81}.

\subsection{Interband spectral range\label{Sec3.1}}

\begin{figure}[tbp]
\includegraphics[width=8.6cm]{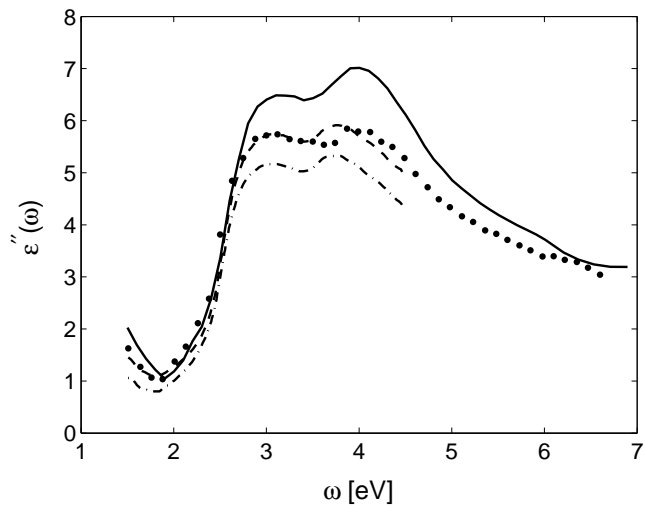}\newline
\caption{The imaginary part of the dielectric function in the
interband region for different samples. The solid line represents
well annealed bulk-like film by Th\`{e}ye \cite{The70}. The dots
are the data by Johnson and Christy \cite{Joh72} found for
unannealed films. The dashed and dash-dotted lines are the modern
data by Wang et al. \cite{Wan98} for unannealed films. They
correspond to the films deposited with e-beam and thermal
evaporation methods, respectively.} \label{fig4}
\end{figure}

Let us start from the interband absorption or domain {\it 3}. This
range is not very significant for the Casimir force but it provides us with
important information how the sample
preparation procedure influences the data. There is significant
amount of data in this range obtained by combined reflectance and
transmittance \cite{The70,Joh72}, ellipsometric spectroscopies
\cite{Pel69,Gue75,Asp80,Ben99,Wan98} on unannealed \cite
{Joh72,Asp80,Ben99,Wan98} or annealed \cite{The70,Pel69,Gue75}
thin film \cite{The70,Joh72,Asp80,Ben99,Wan98} or bulk
\cite{Pel69,Gue75} samples measured in air
\cite{Joh72,Asp80,Ben99,Wan98} or ultrahigh vacuum \cite
{The70,Pel69,Gue75}. A representative picture of typical variations of
$\varepsilon ^{\prime \prime }\left( \omega \right) $ is shown in Fig. \ref
{fig4}. The data for this figure were chosen to represent very dense
(bulk-like) film \cite{The70}, the films similar to those used for the
Casimir force experiments \cite{Wan98}, and some intermediate case \cite
{Joh72}.

Th\`{e}ye \cite{The70} described her films very carefully. The samples
were semitransparent $Au$ films with the thickness $100-250\ $\AA\
evaporated in ultrahigh vacuum on supersmooth fused silica. The deposition
rate was a few \AA $/s$ and the substrate was kept in most cases at room
temperature. After
the deposition the films were annealed under the same vacuum at $%
100-150^{\circ }\ C$, the process being monitored by simultaneous
dc resistance measurements. The structure of the films was
investigated by X-ray and electron-microscopy methods. The
''good'' films were continuous formed of joined regularly shaped
crystallites, the average lateral size of which ranged from 300 to
500 nm. The main defects were grain boundaries. It was
demonstrated that flatness and smoothness of the film surfaces
were very good. The dc resistivity of the films were found to be
very sensitive to the conditions of preparation. The errors in the
optical characteristics of the films were estimated on the level
of a few percents.

Johnson and Christy \cite{Joh72} evaporated their films onto fused quartz
substrates at room temperature. The evaporation rate was relatively high,
60 \AA /s, and the thickness was in the range 250-500 \AA . The presented
data \cite{Joh72} were found for unannealed samples. The structural
characterization of the films was not made. Errors in the optical data were
estimated on the level of 2\%.

Wang et al. \cite{Wan98} evaporated the films on fused silica or optical
glasses with two different methods. For the thermal evaporation the rate was
50-100 \AA /s and for the electron-beam evaporation it was 5 \AA/s.
The substrates were kept at room temperature and there was no annealing
after the deposition. Roughness of the films or structural details were not
reported. Opaque $150\ nm$ thick films were characterized
ellipsometrically
in the photon energy range $1.5-4.5\ eV$. Errors were not specified but
typical precision of the method is of about 1\%.

As one can see in Fig. \ref{fig4} the spectra differ mainly on the
scale factor. The experimental errors are too small to explain
significant variation of the data. One can conclude that there is
genuine sample dependance of optical characteristics of gold films.
Differences among these spectra have been attributed to various
effects such as roughness \cite {The70,Win75}, surface films \cite
{Bea65}, unspecified internal electron-scattering defects \cite
{The70,Jun76}, strain \cite {Win75}, grain-boundary material \cite
{Win75,Hun73,Nag74}, and voids \cite {Hod68}. Special investigation
directed to quantitative understanding the discrepancies on a
systematic basis was undertaken by Aspnes et al. \cite{Asp80} (see
also \cite {Asp95} and \cite{Lyn95}). The authors analyzed the
literature data and their own gold films evaporated by electron beam
or sputter-deposited.
The best films were obtained on cleaved $NaCl$ substrates. Glass, $%
SiO_{2}$, and sapphire substrates were also investigated. Air cleaved
$NaCl$ substrates were superior because gold start to grow on them
epitaxially. Scaling behavior of the spectra was attributed to
different volumes of voids in the films prepared by different
methods. It was also stressed that grains in a film imply voids,
because it is impossible to produce grain boundaries in fcc or
hexagonal close-packed lattices without losing density.

To describe the contribution of voids in the dielectric function,
the Bruggeman effective medium approximation \cite{Bru35} was used
\cite{Asp80}. Nowadays this approximation is successfully applied
for interpretation of the spectra of different nanocomposite
materials (see, for example, \cite {Dal00,Che03}). If $\varepsilon
$ is the dielectric function of $Au$ in its homogeneous form, then
the effective dielectric function, $\left\langle \varepsilon
\right\rangle $, of the material containing a volume fraction of
voids $f_{V}$ can be found from the equation

\begin{equation}
\frac{\left\langle \varepsilon \right\rangle -\varepsilon _{H}}{\left\langle
\varepsilon \right\rangle +2\varepsilon _{H}}=f_{V}\frac{1-\varepsilon
_{H}}{%
1+2\varepsilon _{H}}+(1-f_{V})\frac{\varepsilon -\varepsilon _{H}}{%
\varepsilon +2\varepsilon _{H}},  \label{EMA}
\end{equation}

\noindent where $\varepsilon _{H}$ is the dielectric function of
the ''host'' material. In the Bruggman approximation it is assumed
that the host material coincides with the effective medium,
$\varepsilon _{H}=\left\langle \varepsilon \right\rangle $, so it
treats both void and material phases on an equal. If the void
fraction is small, $f_{V}\ll 1$, then the solution is

\begin{equation}
\left\langle \varepsilon \right\rangle \approx \epsilon \left( 1-
3f_{V}\frac{%
\varepsilon -1}{2\varepsilon +1}\right) .  \label{SolEMA}
\end{equation}

\noindent Aspnes et al. \cite{Asp80} analyzed available data for $%
\varepsilon ^{\prime \prime }\left( \omega \right) $ in the
interband spectral range fitting them with Eq. (\ref{EMA}). It was
found out that in their e-beam evaporated unannealed sample the
volume fraction of voids was 4\% larger than in Th\`{e}ye annealed
thin films \cite{The70} but 5\% smaller than $f_{V}$ in Johnson
and Christy \cite{Joh72} unannealed thin films. One of the films
was sputter-deposited on $NaCl$ substrate that was held at liquid
nitrogen temperature. This sample was full of voids that were
clearly seen with the transmission electron microscopy. The volume
fraction was estimated as $f_{V}=25\%$ in respect to Th\`{e}ye
\cite{The70}
film. One should mention that roughness of the film also will reduce $%
\varepsilon ^{\prime \prime }\left( \omega \right) $ in the
interband region but this is a minor effect \cite{Asp80} because
the films in Refs. \cite {The70,Joh72,Asp80} were quite smooth.
However, significant roughness could be the reason why Wang et al.
\cite{Wan98} thermally evaporated film shows low absorptivity.

Concluding this discussion one can say that the sample preparation
procedure
significantly influences the optical properties of the films in the
interband spectral range. At first sight it seems unimportant for the
Casimir force since variation of the dielectric function in this range gives
only minor correction to the force. However, the volume fraction of voids,
$f_{V}$, that can be estimated from these data defines the exact value of
the effective plasma frequency which is very important for the precise
prediction of the Casimir force. To see it clearly consider Eq.
(\ref{SolEMA}) at low frequencies where $\left| \varepsilon \right| \gg 1$.
In this limit the effective dielectric function will be simply proportional
to that of the homogeneous material

\begin{equation}
\left\langle \varepsilon \right\rangle \approx \epsilon \left( 1-\frac{3}{2}%
f_{V}\right) .  \label{lfEMa}
\end{equation}

\noindent In the Drude region where $\varepsilon $ is given by Eq.
(\ref {Drude}) this relation is equivalent to the existence of
the effective plasma frequency

\begin{equation}
\left\langle \omega _{p}^{2}\right\rangle \approx \omega _{p}^{2}\left(
1-\frac{3}{2}f_{V}\right)  \label{Ompeff}
\end{equation}

\noindent because voids reduce the number of conduction electrons per unit
volume.

Correction to the force due to variation of $\varepsilon ^{\prime
\prime }\left( \omega \right) $ in the interband spectral range is
expected to be small. Nevertheless, let us estimate it. For
numerical calculation it was assumed that the bulk material can be
described by Th\`{e}ye data \cite {The70} but the film material is
given by Johnson and Christy data \cite {Joh72}. Besides, it
was assumed that the difference between these data is negligible
at the photon energies larger than $6.5\ eV$. For the third term
in Eq. (\ref{DF}) and for the sphere-plate geometry it was found
$\left( \Delta F/F\right)_{3}=-0.004,\ -0.002,\ -0.001$ for the separations
$a=50,\ 100,\ 150\ nm$, respectively. Indeed, the correction is
negative since any deviation from the perfect material will reduce
the reflection coefficients and, therefore, the absolute value of
the force will be smaller. At separations larger than $150\ nm$ the
correction to the force is completely negligible on the level of 0.1\%.

\subsection{Intraband spectral range\label{Sec3.2}}

When the photon energy is below the interband threshold, $\omega
<\omega _{0} $, only free carriers should contribute to the
dielectric response. However, the Drude function (\ref{Drude})
obviously fails to describe the data nearby the transition
(see points below $\omega _{0}$ in Fig. \ref{fig4}). Steep rise of
$\varepsilon ^{\prime \prime }$ near $\omega_{0}=2.45\ eV$ originates
from $d$ bands to Fermi-surface transition near $L$ point \cite
{Coo65} but there is a long tail which extends well below $2\ eV$. This
tail cannot be explained by the temperature broadening of the threshold. It
was demonstrated \cite{Gue75} that an additional transition at $1.94\ eV$
near $X $ point in the Brillouin zone is involved. This transition has a
smooth edge due to topological reason. In addition, the surface roughness and the
grains can contribute to the tail broadening \cite{Asp80}. This tail is not
very important for the Casimir force but it marks the region where the Drude
parameters cannot be determined precisely.

\begin{figure}[tbp]
\includegraphics[width=8.6cm]{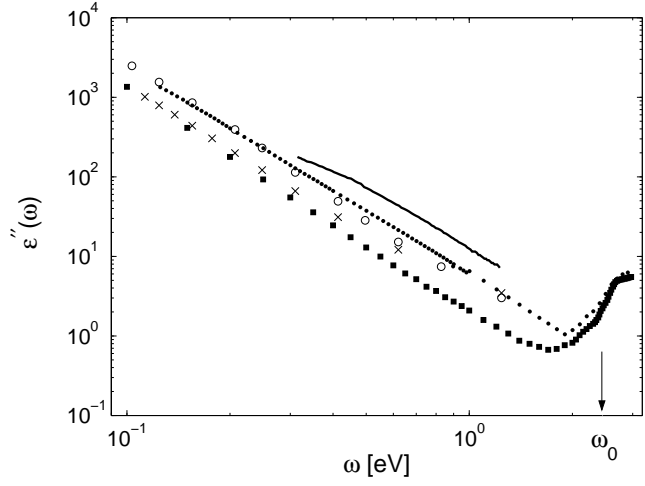}\newline
\caption{Available optical data in the infrared region. The dots
represent the Dold and Mecke data at $\omega<1\ eV$ for evaporated
unannealed films \cite{Dol65} and Th\`{e}ye data \cite{The70} for
higher frequencies. The squares are the data \cite{Wea81} for
electropolished bulk $Au(110)$. The open circles are the data for
carefully prepared well annealed evaporated films \cite{Mot64}. The
crosses represent the data \cite{Pad61} for unannealed evaporated films.
The only modern data \cite{Xia00} are given by the solid line.}
\label{fig5}
\end{figure}

In contrast with the interband region there are only a few sources where the
dielectric function was measured in the infrared. The available data for $%
\varepsilon ^{\prime \prime }\left( \omega \right) $ in the range
$\omega <\omega _{0}$ are presented in Fig. \ref{fig5}. The dots
represent the points from the handbook \cite{HB1} comprising Dold
and Mecke data \cite {Dol65} for $\omega <1\ eV$ and Th\`{e}ye
data \cite{The70} for higher frequencies. Dold and Mecke did not
describe carefully the sample preparation. It was reported only
\cite{Dol65} that the films were evaporated onto a polished glass
substrate and measured in air by using an ellipsometric technique.
Annealing of the samples was not reported. These data are typically
referred in connection with the Casimir force calculation. The
squares are the data by Weaver et al. \cite{Wea81} that were
found for the electropolished bulk \ $Au(110)$ sample.
Originally the reflectance was measured in a broad interval
$0.1\leq \omega \leq 30\ eV$ and then the dielectric function was
determined by the Kramers-Kronig analysis. Due to indirect
determination of \ $\varepsilon $ the recommended accuracy of
these data is only 10\%. Motulevich and Shubin \cite{Mot64} result
for $Au$ films is marked with the open circles. In this paper the
films were carefully described. Gold was evaporated on polished
glass at pressure $\sim 10^{-6}\ torr$. The investigated films
were $0.5-1\ \mu m$ thick. The samples were annealed in the same
vacuum at $400^{\circ }\,C$ fore more than 3 hours. The optical
constants $n$ and $k$ ($n+ik=\sqrt{\varepsilon }$) were measured
by the polarization methods in the spectral range $1-12\ \mu m$.
The errors in $n$ and $k$ were estimated as 2-3\% and 0.5-1\%,
respectively. Moreover, the density and conductivity of the
films were measured. The crosses represent Padalka and
Shklarevskii data \cite{Pad61} for evaporated unannealed $Au$ films.
The solid line shows the only modern data \cite{Xia00} for
$Au$ films sputtered on $Si$ substrate. No additional details
about the film preparation were reported. The precision of optical
data was claimed to be better than 1\% in the entire spectral
range.

Different researches stressed that the conduction electrons are
much more sensitive to slight changes in the material structure
\cite {The70,Asp80,Ben65}. As one can see in Fig. \ref{fig5} it is,
indeed, the case. Variety of the data in any case cannot be
explained by the experimental errors. The observed scattering of
the data is the result of different preparation procedures and
reflects genuine difference between samples. Bennett and Ashley
\cite{Ben65} observed considerable reduction in the reflectance of
$Au$ films when instead of ultrahigh vacuum ($5\cdot
10^{-9}\ torr$) the films were evaporated in the ordinary vacuum ($%
10^{-5}\ torr$). It was noted also that it was more difficult to
prepare films with reproducible reflectance in the ordinary
vacuum. Even more important are
the deposition method (there is a whole spectrum of evaporation
and sputtering methods), type of the substrate, its temperature
and quality, and deposition rate. When we are speaking about
the Casimir force measurement with the precision of 1\% or
better there is no any more such a material as gold in general.
There is only a material prepared in definite conditions.

The data in Fig. \ref{fig5} below $\omega _{0}$ should be
described quite well with the Drude function

\begin{equation}
\varepsilon ^{\prime \prime }\left( \omega \right) =\frac{\omega
_{p}^{2}\omega _{\tau }}{\omega \left( \omega ^{2}+\omega _{\tau
}^{2}\right) }.  \label{ImDrude}
\end{equation}

\noindent While $\omega \gg \omega _{\tau }$ the data on the
log-log plot should look like a straight line with the slope $-3$
shifted along the ordinate axis mostly due to variation of $\omega _{\tau
}$ for different samples. The data in Fig. \ref{fig5} are in
general agreement with the expectations excluding the solid line.
The latter cannot be described with the Drude behavior, probably,
due to poor preparation of the film. Padalka and Shklarevskii data (crosses)
demonstrate a long Drude tail that can be observed up to
$0.3\ eV$. To all appearance it is again the result of poor sample
preparation. The Drude
parameters can be found fitting both $\varepsilon ^{\prime }$ and $%
\varepsilon ^{\prime \prime }$ with the function (\ref{Drude}).
This procedure will be discussed later.

Now let us estimate the change in the Casimir force due to variation of
$\varepsilon ^{\prime \prime }\left( \omega \right) $ in the infrared. For
this purpose we calculate $\Delta _{n2}$ in Eq. (\ref{deli}) using as $%
\Delta \varepsilon ^{\prime \prime }\left( \omega \right) $ the difference
between the squares \cite{Wea81} and points \cite{HB1} in Fig.
\ref{fig5} in
the interval $\omega _{c}<\omega <\omega _{0}$. With this $\Delta_{n2}$ the
second term in Eq. (\ref{DF}) is calculated that gives the relative change in
the Casimir force due to variation of $\varepsilon ^{\prime \prime }\left(
\omega \right) $ in the indicated frequency interval. For the sphere-plate
geometry it was found that $\left( \Delta F/F\right) _{2}=-0.025,\
-0.020,\ -0.016$ for the separations $a=50,\ 100,\ 150\ nm$, respectively.
The change in the force is
negative and quite large. The main reason for this is the
difference between the relaxation frequencies of the bulk material
(squares) and poorly evaporated film (dots). As one can see in
Fig. \ref{fig5} below $1\ eV$ the dielectric function $\varepsilon
^{\prime \prime }$\ for the film is roughly two times larger than
that for the bulk material (squares) so $\Delta _{n2}$ is
approximately $-0.5$ for all important $n$. There is no
contradiction with the expectation that the force must be larger for
the bulk material. We took
into account here only the frequency interval $\omega _{c}<\omega
<\omega _{0}$. When the contribution of the low-frequency domain
$0<\omega <\omega _{c}$ will be added the total
force will be larger for bulk gold.

\subsection{The Drude parameters\label{Sec3.3}}

The dielectric function in the low-frequency domain, $\omega <\omega _{c}$,
where direct optical data are inaccessible, is defined by the dielectric
function
in the infrared domain, $\omega _{c}<\omega <\omega _{0}$, because
one has to
extrapolate the optical data from this domain to inaccessible region. The
real and imaginary parts of $\varepsilon $ follow from the Drude
function (\ref{Drude}) with an additional term ${\cal P}$ in $\varepsilon
^{\prime }$:

\begin{equation}
\varepsilon ^{\prime }\left( \omega \right) ={\cal P}-\frac{\omega
_{p}^{2}}{\omega ^{2}+\omega _{\tau }^{2}},\quad \varepsilon
^{\prime \prime }\left(\omega \right) =\frac{\omega _{p}^{2}\omega
_{\tau }}{\omega \left( \omega^{2}+\omega _{\tau }^{2}\right) }.
\label{DrudeRI}
\end{equation}

\noindent The polarization term ${\cal P}$ has appeared here due to the
following reason. The total dielectric function $\varepsilon
=\varepsilon _{\left( c\right) }+\varepsilon _{\left( i\right) }$
includes contributions due to conduction electrons $\varepsilon
_{\left( c\right) }$ and the interband transitions $\varepsilon
_{\left( i\right) }$. The polarization term consists of the atomic
polarizability and polarization due to the interband transitions $
\varepsilon _{\left( i\right) }^{\prime }$

\begin{equation}
{\cal P}=1+\frac{N_{a}\alpha }{\varepsilon _{0}}+\varepsilon _{\left(
i\right) }^{\prime }\left( \omega \right) ,  \label{polariz}
\end{equation}

\noindent where $\alpha $ is the atomic polarizability and $N_{a}$ is the
concentration of atoms. If there is no the Drude tail the last term can be
considered as a constant. This is because the interband transitions have
the threshold behavior with the onset $\omega _{0}$ and the Kramers-Kronig
relation allows one to express $\varepsilon _{\left( i\right) }^{\prime }$ as

\begin{equation}
\varepsilon _{\left( i\right) }^{\prime }\left( \omega \right) =\frac{2}{\pi
}\int\limits_{\omega _{0}}^{\infty }dx\frac{x\varepsilon _{\left( i\right)
}^{\prime \prime }\left( x\right) }{x^{2}-\omega ^{2}}.  \label{KKi}
\end{equation}

\noindent For $\omega \ll \omega _{0}$ this integral does
not depend on $\omega $. In reality the situation is more complicated
because the transition is not sharp and many factors can
influence the structure of the Drude tail. It will be assumed here that
${\cal P}$ is a constant but the fitting procedure will be shifted to
frequencies where the Drude tail is not important. In practice Eq.
(\ref{DrudeRI}) can be applied for $\omega <1\ eV$.

As we have seen above the plasma frequency $\omega _{p}$ of a
sample under investigation can deviate from that of perfect bulk
material due to presence of voids (see Eq. (\ref{Ompeff})). The
Casimir force is measured between films which are far from
perfect. The best optical properties demonstrate well annealed
films prepared in ultrahigh vacuum \cite{The70,Ben65}. The
annealed films were not used for the force measurement and it is
hardly reasonable to do in any future experiments \footnote{
Annealing leads to a very specific kind of roughness: grooving.
Convincing AFM images of annealed medium thick gold films one can
see, for example, in \protect{Ref. \cite{Liu97}}. The grooving is
not important for the infrared optical properties but it
significantly complicates calculation of the correction to the
force due to the surface roughness.}. For high precision
calculation of the force the procedure proposed in Ref.
\cite{Lam00} is not good any more. One cannot use the bulk plasma
frequency to extrapolate the dielectric function to experimentally
inaccessible frequencies. The second Drude parameter, the
relaxation frequency $\omega _{\tau }$, is even more sensitive to
the sample preparation procedure. Therefore, the Drude parameters
have to be extracted from the optical characteristics of the same
sample which is used for the force measurement. Our purpose here
is to establish the magnitude of the force change due to
reasonable variation of the optical properties. To this end the
available low-frequency data for $\varepsilon ^{\prime \prime
}\left( \omega \right) $ shown in Fig. \ref{fig5} and $\varepsilon
^{\prime }\left( \omega \right)$ (not shown) were fitted with
Eq. (\ref{DrudeRI}). The results together with the expected errors are
collected in Table \ref{tab1}.

\begin{table}
\centering
\begin{tabular}{l||l|l|l|l}
N & $\omega _{p},\ (eV)$ & $\omega _{\tau }\cdot 10^{2},\ (eV)$ &
${\cal P}$ & Ref., mark \\ \hline\hline 1 & $7.52\pm 0.21$ &
$6.1\pm 0.9$ & $-28\pm 67$ & \cite{Dol65,HB1}, $\ \cdot$ \\
2 & $8.41\pm 0.01\pm 0.42$ & $2.0\pm 0.04\pm 0.3$ & $7.3\pm 6.0$ &
\cite{Wea81}, \ $\blacksquare $ \\
3 & $8.79\pm 0.12\pm 0.09$ & $4.2\pm 0.3\pm 0.2$ & $14\pm 79$ &
\cite{Mot64}, \ $\bigcirc $ \\
4 & $6.81\pm 0.07$ & $3.6\pm 0.3$ & $-21\pm 39$ & \cite{Pad61}, \ X
\\
5 & $9.04\pm 0.03$ & $2.67\pm 0.03$ & $24\pm 117$ & \cite{Ben66}
\end{tabular}
\caption{The Drude parameters found by fitting the available
infrared data for $\varepsilon ^{\prime }\left( \omega \right)$
and $\varepsilon ^{\prime \prime }\left( \omega \right) $ with Eq.
(\ref{DrudeRI}). The first error is the statistic one
 and the second error (if present) is the systematic one.}\label{tab1}
\end{table}

The first error in Table \ref{tab1} is the statistical one. It was
found using $\chi ^{2}$ criterion for joint estimation of 3
parameters \cite{PatDat}. The second error is the systematic one.
It was found propagating the optical errors, when indicated, to
$\omega _{p}$ and $\omega _{\tau }$. One can see that the
polarization term ${\cal P}$ cannot be resolved in the frequency
range where $\varepsilon ^{\prime }$ is very large. Significant
variation of the plasma frequency, well above the errors, is a
distinctive feature of the table. Carefully prepared annealed
samples demonstrate larger value of $\omega _{p}$. The rows 1 and
4 corresponding to the evaporated unannealed films have
considerably smaller $\omega _{p}$. Note that our calculations are
in agreement with that given by the authors \cite{Dol65,Pad61}
themselves. To all appearance these low values of $\omega _{p}$
are the result of poor preparation of the films. In accordance
with the effective medium approximation at low frequencies the
effective plasma frequency is given by Eq. (\ref{Ompeff}). Using
this equation and $\omega _{p}=9.0\ eV$ for homogeneous gold the
fraction of voids $f_{V}=0.20$ and $0.28$ was found for the rows 1
and 4, respectively. Discussing the interband transitions we have
seen that this large fraction of voids is possible for poor
prepared films like that sputtered at liquid nitrogen temperature
\cite{Asp80} or thermally evaporated \cite{Wan98}.

The fifth row in Table \ref{tab1} should be discussed separately.
It is based on the reflection data \cite{Ben65} in the infrared
for carefully prepared opaque films evaporated and annealed in
ultrahigh vacuum on extremely smooth fuzed quartz substrate. The
films have the highest reflectivity in the infrared of that ever
reported. These data have been analyzed by Bennett and Bennett
\cite{Ben66} using the conductivity of bulk material and assuming
that there is one conduction electron per atom. Perfect agreement
with the Drude theory in the wavelength range $3-32\ \mu m$ was
found with no adjustable parameters. The errors in the row 5
demonstrate rather the level of this agreement than the
statistical or systematic errors.

\begin{figure}[tbp]
\includegraphics[width=8.6cm]{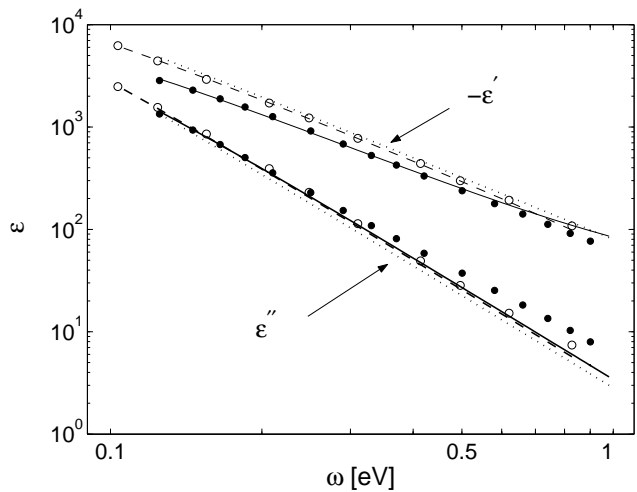}\newline
\caption{The infrared optical data by Dold and Mecke \cite{Dol65}
(dots) and by Motulevich and Shubin \cite{Mot64} (open circles)
together with the best Drude fits given by the solid and dashed lines,
respectively. The dotted lines present the fit which is typically
used in the calculation of the Casimir force \cite{Lam00}. The
latter one much better agrees with the open circles than with the
handbook data (dots). } \label{fig6}
\end{figure}

To have an idea how good is the fitting procedure, in Fig.
\ref{fig6} the experimental points and the best fitting curves are
shown for Dold and Mecke data \cite{Dol65} (dots and solid lines)
and Motulevich and Shubin data \cite{Mot64} (open circles and
dashed lines). One can see that for $\varepsilon ^{\prime \prime
}$ at high frequencies the dots lie above the solid line
indicating the presence of the Drude tail. Coincidence of the
solid and dashed lines for $\varepsilon ^{\prime \prime }$ is
accidental. The fits for $\varepsilon ^{\prime }$ are nearly
perfect for both sets of the data.

Calculating the Casimir force Lambrecht and Reynaud \cite{Lam00}
assumed that the plasma frequency can be taken for bulk $Au$ with
one conduction electron per atom and found $\omega _{p}=9.0\ eV$. In
contrast with \cite{Ben66} this assumption was not supported by any
optical data. The relaxation frequency $\omega _{\tau }=0.035\ eV$
was found then by fitting Dold and Mecke data \cite{Dol65} for
$\varepsilon ^{\prime \prime }$. The curves corresponding to this
set of parameters are shown in Fig. \ref{fig6} as dotted lines. It
is clear that there is a significant mismatch for $\varepsilon
^{\prime }$. The dotted lines much better agree with Motulevich and
Shubin data \cite{Mot64} than with the data they should describe. If
the force is calculated with the handbook data, the extrapolation to
low frequencies has to be done with the parameters which are in
agreement with the optical data. These parameters are given in the
row 1 (Table \ref{tab1}).

Let us estimate how sensitive is the force to the variation of the Drude
parameters. To this end the first term in Eq. (\ref{DF}) is estimated. In
the domain of interest, $0<\omega <\omega _{c}$, we calculate $\Delta
\varepsilon ^{\prime \prime }\left( \omega \right) $ in Eq. (\ref{deli}) as
the difference between the function (\ref{ImDrude}) with $\omega
_{p}=7.52\ eV,\ \omega _{\tau }=0.061\ eV$ (evaporated film, first row
in Table \ref{tab1}) and the same function with the bulk-like parameters $
\omega _{p}=9.0\ eV,\ \omega _{\tau }=0.035\ eV$ \cite{Lam00}. The
result is the following: $\left( \Delta F/F\right) _{1}=-0.070,\ -0.063,\ -
0.054$ for the separations $a=50,\ 100,\ 150\ nm$, respectively. The
correction is negative since the evaporated film has smaller reflectivity in
comparison with the bulk material. Its magnitude is large because the
force is more sensitive to the behavior of $\varepsilon ^{\prime \prime
}\left( \omega \right) $ in the far infrared than in the infrared or visible
range as was discussed in Sec. \ref{Sec2}. At any circumstances this large
correction cannot be ignored.

The main problem with precise calculation of the Casimir force is to
find the correct values of the Drude parameters to extrapolate the
dielectric function to low frequencies. Obviously, using the bulk
material parameters we
overestimate the force which is measured between the deposited
films. Discussion in this section demonstrates that
the film quality is improved after annealing.
Unannealed films used for the force measurement will inevitably
contain some fraction of voids independently of the used
equipment. This fraction can be larger than 10\% as for the
thermally evaporated films \cite{Wan98} but, probably, not smaller
than 4\% as for the films evaporated with e-beam on $NaCl$
substrate \cite{Asp80} or sputtered at room temperature
\cite{Asp80,Ben99}. The void fraction will influence the effective
plasma frequency. The relaxation frequency is even more sensitive to
the film quality because of electron scattering on the grain
boundaries \cite{Sot03} and other defects. To be able to predict
the force with the precision better than 1\% the films have to be
carefully characterized.

At the end of this section let us estimate the contribution of the
Drude tail to the force. It is calculated in the frequency range
$1\ eV<\omega <\omega _{0}$ that was not used in the
determination of the Drude parameters. We calculate $\Delta \varepsilon
^{\prime \prime }\left( \omega \right) $ in Eq. (\ref {deli}) as
the difference between $\varepsilon^{\prime \prime } $ given by
Eq. (\ref{ImDrude}) with the parameters in the row 1 (Table \ref{tab1})
and $\varepsilon^{\prime \prime } $ from the handbook data.
For the contribution to the force between sphere and plate it was
found $\left( \Delta F/F\right) _{2}=-0.003,\ -0.002,\ -0.001$ for
the separations $a=50,\ 100,\ 150\ nm$, respectively. The effect is
minor.

\section{Upper limit on the Casimir force\label{Sec4}}

While the Drude parameters of the films are not fixed precisely from
the optical characterization, one
can establish a reliable upper limit on the force. It is based on the
observation \cite{Sve00a,Sve00b} that the reflection coefficients $\left|
r_{s,p}\right| $ in the Lifshitz formula (\ref{Fpp}) or (\ref{Fsp}) are
always larger for perfect material than for any real
material containing different kind of defects. As the result the force
between perfect bodies will be the largest. To be more precise, at a
fixed imaginary frequency $\left| r_{s}\right| $ and $\left| r_{p}\right| $
are monotone increasing functions of $\varepsilon \left( i\zeta \right) $.
The dielectric function $\varepsilon \left( i\zeta \right) $ can be
represented in the following form:

\begin{equation}
\varepsilon \left( i\zeta \right)-1 =\frac{\omega _{p}^{2}}{\zeta \left(
\zeta +\omega _{\tau }\right) }+\frac{2}{\pi }\int\limits_{\omega
_{0}}^{\infty }d\omega \frac{\omega \left[ \varepsilon ^{\prime \prime
}\left( \omega \right) -\varepsilon _{D}^{\prime \prime }\left( \omega
\right) \right] }{\omega ^{2}+\zeta ^{2}}+\frac{2}{\pi }\int\limits_{\omega
_{1}}^{\omega_{0} }d\omega \frac{\omega \left[ \varepsilon ^{\prime \prime
}\left( \omega \right) -\varepsilon _{D}^{\prime \prime }\left( \omega
\right) \right] }{\omega ^{2}+\zeta ^{2}},  \label{epsmax}
\end{equation}

\noindent where $\varepsilon _{D}^{\prime \prime }\left( \omega
\right) $ is the Drude function (\ref{ImDrude}). The first
two terms in Eq. (\ref {epsmax}) will be maximal for a
single-crystal when $\omega_{p} $ is the largest, $\omega_{\tau} $
is the smallest, and the interband ($\omega>\omega_{0} $)
absorption is maximal. The same is not true in the range of the
Drude tail, $\omega _{1}<\omega <\omega _{0}$. As one can see in
Fig. \ref{fig5} the smallest Drude tail shows the single crystalline
sample, while the films demonstrate much larger tail. However, the
increase of $\varepsilon \left( i\zeta \right) $ due to the
tail is always small in comparison with the decrease of $%
\varepsilon \left( i\zeta \right) $ due to deviation of the Drude parameters
from the bulk values. Hence, $\varepsilon \left( i\zeta \right) $ will be the
largest for single crystalline gold.

It would be natural to take the dielectric function of the
single crystalline sample \cite{Wea81} to find the upper limit
on the Casimir force. This possibility was rejected due to the
following reason. Precision of these data is only 10\% because the
Kramers-Kronig analysis was used to get both of the optical
constants. This is the reason for large systematic errors in the
Drude parameters (see row 2 in Table \ref{tab1}). Since the
force is very sensitive to the value of $\omega _{p}$, the
precision of the data is not good for our purpose. To get the
Drude parameters of the perfect sample, the plasma frequency was
calculated with the help of Eq. (\ref{Omp}) assuming that each atom
gives one conduction electron with the effective mass $m_{e}^{\ast }=m_{e}$.
In this way it was found $\omega _{p}=9.03\ eV$. The
relaxation frequency was calculated then from the resistivity of bulk gold
at room temperature $\rho =\omega _{\tau }/\varepsilon _{0}\omega
_{p}^{2}=2.25\ \mu \Omega \cdot cm$ that gave $\omega _{\tau}=0.025\ eV$.
The deviation from the parameters presented in the row 2 is a little bit
larger than one standard deviation but acceptable. Note that the
parameters found in this way agree very well with Bennett and Bennett
data \cite{Ben66} (row 5 in Table \ref{tab1}).

\begin{figure}[tbp]
\includegraphics[width=8.6cm]{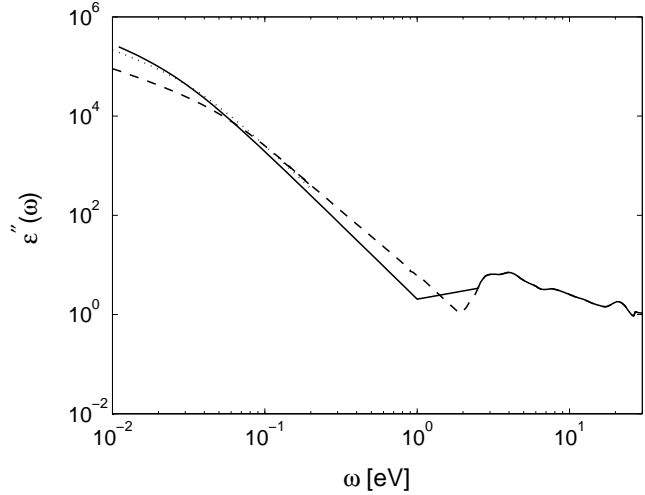}\newline
\caption{The dielectric function used to calculate the upper limit
on the Casimir force (solid line). The dashed line represents the
handbook data \cite{HB1} extrapolated to low frequencies
$\omega<0.125\ eV$ continuously for both $\varepsilon ^{\prime }$
and $\varepsilon ^{\prime \prime }$. The dotted line shows the
extrapolation used in Ref. \cite{Lam00} which is continuous only
for $\varepsilon ^{\prime \prime }$.} \label{fig7}
\end{figure}

The upper limit on the force is calculated using the Drude function
(\ref{ImDrude}) at $\omega<\omega _{1}=1\ eV$ with the
parameters $\omega _{p}=9.03\ eV$, $\omega _{\tau }=0.025\ eV$.
Starting from $\omega =\omega _{0}=2.45\ eV$ \ Th\`{e}ye data
\cite{The70,HB1} are used as the best data known in the interband
region. At frequencies $\omega>6\ eV$ the handbook data are used. The
force is not sensitive to the reasonable variation of the data due to the
sample effect at $\omega >6\ eV$. The most problematic interval is the
range of the Drude tail, $\omega_{1}<\omega <\omega _{0}$. In this
interval the data are interpolated
linearly on log-log plot between $\varepsilon ^{\prime \prime }\left(
\omega _{1}\right) $ and $\varepsilon ^{\prime \prime }\left( \omega
_{0}\right) $. Of course, any real data should lie below the interpolated
line. The only justification for this procedure is the fact that we are
trying to find the upper limit on the force. This interpolation gives the
largest contribution. If, for example, the Drude function is
continued up to $\omega =\omega _{0}$ then the relative change in the
force will be $\Delta F/F=-0.005,\ -0.003,\ -0.002$ for
$a=50,\ 100,\ 150\ nm$, respectively. The function $\varepsilon ^{\prime
\prime }\left( \omega \right) $ used for the calculation of the upper limit is
shown in Fig. \ref{fig7} by the solid line. Note that it is not just an
abstract line for the best gold sample. In the interval $0.039<\omega
<0.41\ eV$ it is very close to the data for the films carefully prepared in
ultrahigh vacuum \cite{Ben66}.
The handbook data presented in the same figure by the dashed line for $
\omega >0.125\ eV$. Two extrapolations to low frequencies are also
shown. The dashed line at $\omega <0.125\ eV$ corresponds to the best
fit of the data in the infrared range $0.125<\omega <1\ eV$ providing
smooth continuation of both $\varepsilon ^{\prime }\left( \omega \right) $
and $\varepsilon ^{\prime \prime }\left( \omega \right) $ to lower
frequencies. The dotted line gives the extrapolation used in Ref.
\cite{Lam00} with the parameters $\omega _{p}=9.0\ eV$, $\omega
_{\tau }=0.035\ eV$. This extrapolation is continuous only for
$\varepsilon ^{\prime \prime }\left( \omega \right) $. Mismatch of
$\varepsilon ^{\prime }$ at $\omega =0.125\ eV$ is 100\% as one can
see in Fig. \ref{fig6} (dotted line).

One can ask why in the infrared the solid line representing the best
sample lies below the handbook data. Larger $\varepsilon ^{\prime \prime
}$ seems augments $\varepsilon \left( i\zeta \right) $ and, therefore,
increases the force. In reality the integral effect is important.
Larger values of $\varepsilon ^{\prime \prime }$
in the infrared happen mostly because of larger $\omega _{\tau }$ (see Eq.
(\ref{ImDrude})). In the far infrared the function with larger
$\omega _{\tau }$ will be smaller than that for the best sample.
Because the contribution of low frequencies is more important in the
integral (\ref{K-K}), $\varepsilon \left( i\zeta \right) $ always
will be larger for the perfect single-crystal.

The results of calculation of the upper limit on the Casimir force are
presented in Table \ref{tab2}. It gives the so-called reduction factors
\cite{Lam00} which are defined as

\begin{equation}
\eta _{pp}=\frac{F_{pp}\left( a\right) }{F_{pp}^{c}\left( a\right) },\quad
F_{pp}^{c}\left( a\right) =-\frac{\pi ^{2}}{240}\frac{\hbar c}{a^{4}}
\label{redpp}
\end{equation}

\noindent for the plate-plate geometry and

\begin{equation}
\eta _{sp}=\frac{F_{sp}\left( a\right) }{F_{sp}^{c}\left( a\right) },\quad
F_{sp}^{c}\left( a\right) =-\frac{\pi ^{3}R}{360}\frac{\hbar c}{a^{3}}
\label{redsp}
\end{equation}

\noindent for the sphere of radius $R$ and plate. Here $F_{pp}$
and $F_{sp}$ are the forces given by the Lifshitz formulas
(\ref{Fpp}) and (\ref{Fsp}), respectively, and $F_{pp}^{c}$,
$F_{sp}^{c}$ are the corresponding expressions for the Casimir
force between bodies made of the ideal metal. The reduction factors
are given for the case when the temperature correction is negligible.
It corresponds to the plasma model prescription
\cite{Bor00} for the $n=0$ term in the Lifshitz formula.
Corrections to the reduction factors are also shown for the cases
when the zero term is described by the Drude model $\Delta \eta
^{BS}$ \cite{Bos00a} or coincides with the ideal metal $\Delta
\eta ^{SL}$ \cite{Sve01}. They give the temperature
corrections in different approaches.

\begin{table}
\centering
\begin{tabular}{l||l|l|l||l|l|l}
$a\ (nm)$ & $\eta _{pp}$ & $\Delta \eta _{pp}^{BS}$ & $\Delta \eta
_{pp}^{SL}$
& $\eta _{sp}$ & $\Delta \eta _{sp}^{BS}$ & $\Delta \eta _{sp}^{SL}$
\\
\hline\hline
$60$ & 0.364 & -0.002 & 0.007 & 0.435 & -0.005 & 0.009 \\
$80$ & 0.424 & -0.003 & 0.009 & 0.497 & -0.008 & 0.011 \\
$100$ & 0.472 & -0.005 & 0.010 & 0.546 & -0.011 & 0.012 \\
$120$ & 0.513 & -0.007 & 0.011 & 0.586 & -0.015 & 0.013 \\
$140$ & 0.547 & -0.010 & 0.012 & 0.620 & -0.019 & 0.013 \\
$160$ & 0.578 & -0.012 & 0.013 & 0.648 & -0.023 & 0.014 \\
$180$ & 0.604 & -0.014 & 0.013 & 0.673 & -0.027 & 0.015 \\
$200$ & 0.627 & -0.017 & 0.014 & 0.694 & -0.031 & 0.015 \\
$220$ & 0.648 & -0.020 & 0.014 & 0.713 & -0.035 & 0.015 \\
$240$ & 0.667 & -0.022 & 0.014 & 0.730 & -0.039 & 0.016 \\
$260$ & 0.684 & -0.025 & 0.015 & 0.745 & -0.044 & 0.016 \\
$280$ & 0.699 & -0.028 & 0.015 & 0.759 & -0.048 & 0.016 \\
$300$ & 0.713 & -0.030 & 0.015 & 0.771 & -0.052 & 0.016
\end{tabular}
\caption{The reduction factors for the plate-plate geometry
$\eta_{pp}$ and sphere-plate geometry $\eta_{sp}$ corresponding to
the upper limit on the Casimir force. $\Delta \eta^{BS}$ gives
the temperature correction according to Ref. \cite{Bos00a} and
$\Delta \eta^{SL}$ gives the same correction according to Ref.
\cite{Sve01}.}\label{tab2}
\end{table}

The values presented in Table \ref{tab2} one can compare with the result by
Lambrecht and Reynaud \cite{Lam00} who gave for $a=100\ nm$ the
values $\eta_{pp}=0.48$ and $\eta _{sp}=0.55$. They are very close to
that presented in Table \ref{tab2}. This coincidence is because the
increase of the force due to the infrared region was compensated by
its decrease in the far infrared since larger value
$\omega _{\tau }=0.035\ eV$ was used.
Therefore, the result of Ref. \cite{Lam00} represents rather the upper
limit on the force than the real force corresponding to the handbook
data. The latter force should be calculated with the Drude parameters
given in Table \ref{tab1} (row 1). For the sphere-plate geometry
at $a=100\ nm $ it is 6.3\% smaller as was demonstrated in
Sec. \ref{Sec3.3}. For small
separations presented in Table \ref{tab2} $\Delta \eta /\eta $ gives the
relative temperature correction. It is well known that this correction is
negative for the Bostr\"{o}m and Sernelius approach \cite{Bos00a}
making the absolute value of the force smaller; it is positive when the
$n=0$ term is assumed to coincide with that for the ideal metal
\cite{Sve01}. In the latter case the relative correction is nearly constant at
small separations and is of about 2\% independently on the geometry. In
the former case it decreases from -0.5\% to -4.2\% for the plate-plate and
from -1.1\% to -6.7\% for sphere-plate geometries.

We did not include in the upper limit the correction due to
surface roughness \cite{Kli96,Gen03} for obvious reason: it should
be specified for a definite experiment. This correction always
increases the absolute value of the force. Additionally, the
nonlocal effects in the interaction of electromagnetic field with
a solid were not included in the upper limit on the force. It was
demonstrated \cite{Esq03} that the nonlocal screening effect due
to excitation of plasma oscillations in metals will reduce the
Casimir force on a few percents at small separations. The
influence of the anomalous skin effect on the force was also
analyzed recently \cite{Esq04}. It was found that this effect
gives only minor correction which reduces the force smaller that
0.5\%.

\section{Expected magnitude of the sample effect \label{Sec5}}

As we have seen above the main uncertainty of the Casimir force
comes from the uncertainty of the Drude parameters. The reason for
variation of the effective plasma frequency, $\omega _{p}$, is the
presence of voids in the deposited films as was discussed in Sec.
\ref{Sec3.1}. The relaxation frequency, $\omega _{\tau }$, is much
more sensitive to the details of the film preparation
\cite{The70,Asp80,Ben65}. It is well known that the resistivity
$\rho $\ of deposited films is larger than that for the single
crystalline material. The resistivity of unannealed medium thick
($150-200\ nm $) gold films \cite{Cha84} was ranged from 4 to
$10\ \mu\Omega \cdot cm$ in dependence on the method (electron beam,
sputtering) and rate (1 \AA /s, 20 \AA /s) of deposition on
$Si$-substrate. For well crystallized annealed film deposited in
ultrahigh vacuum Th\`{e}ye \cite{The70} has found $\rho =2.97\ \mu
\Omega \cdot cm$ but for unannealed film this value was as high as
$4.40\ \mu \Omega \cdot cm$. These values should be compared with
the resistivity of bulk gold $2.25\ \mu \Omega \cdot cm$. The
difference between film and bulk values was attributed to the
presence of grain boundaries in the films \cite{The70}. In this section a
simple two parameter model will be presented allowing one to estimate
the force between films via the volume fraction of voids $f_{V}$ and
the mean grain size $D$.

\subsection{Optical conductivity as a function of the grain size
\label{Sec5.1}}

Mayadas and Shatzkes \cite{May70} proposed the first model for
calculation of the dc conductivity, $\sigma=1/\rho $, of films
taking into account the electron scattering on the grain
boundaries. This model successfully describes films and nanowires
\cite{Rei86,Dur00,Ste02}. In the model the films were treated as
joined grains with bulk material properties; the grain boundaries
contributed to the electron transport. A semiclassical approach
based on the Boltzmann transport equation was used with the
collision term associated to the grain boundaries. The background
scattering was taken into account via the bulk relaxation time.
The grains were considered to have columnar shape to be laterally
bounded by planes. The grain boundaries were represented by two
uncorrelated sets of $\delta $-potentials along each lateral
direction. The following expression was found for the conductivity
of films

\begin{equation}
\sigma =\sigma _{0}\left( 1-3\alpha _{0}\left[ \frac{1}{2}-\alpha
_{0}+\alpha _{0}^{2}\ln \left( 1+1/\alpha _{0}\right) \right] \right) ,
\label{cond}
\end{equation}

\noindent where $\sigma _{0}$ is the conductivity of single crystalline
material and the parameter $\alpha _{0}$ is defined as

\begin{equation}
\alpha _{0}=\frac{v_{F}}{D\omega _{\tau }}\frac{{\cal R}}{1-{\cal
R}}.
\label{alpha}
\end{equation}

\noindent Here $v_{F}$ is the Fermi velocity in the homogeneous
material, $D$ is the mean grain size, and ${\cal R}$ is the
reflection coefficient of electrons on the grain boundary. It is
important to stress that $\sigma $ depends only on the mean grain
size; dependence on the standard deviation from the mean value can
be neglected \cite{May70,Sot03}. Variation of the dc conductivity
one can connect with the effective relaxation frequency, $\omega
_{\tau }^{eff}$, by the relation

\begin{equation}
\sigma =\frac{\varepsilon _{0}\omega _{p}^{2}}{\omega _{\tau
}^{eff}}.
\label{cond1}
\end{equation}

\noindent Comparing it with Eq. (\ref{cond}) one can find $\omega _{\tau }^{eff}$.
However, this simple concept does not work at frequencies where the conductivity
demonstrates the dispersion.

Recently the same approach was generalized to the case of optical
conductivity $\sigma \left( \omega \right) $ \cite{Sot03}, which
is converging to the dc conductivity in the low-frequency limit,
and the empirical parameter ${\cal R}$ was extracted from comparison of
the model with the measured reflectivity of gold films of
different grain size. To find the optical conductivity the authors
\cite{Sot03} solved a linearized Boltzmann transport equation
taking into account the anomalous skin effect as well. So, in
general, the conductivity is nonlocal. It means that it depends
not only on frequency but also on the wave vector. It was found
that the anomalous skin effect gives only very small correction to
the reflectance of the films. To simplify the model we will
neglect the nonlocal effect. Recently the anomalous skin effect
was discussed in connection with the Casimir force \cite{Esq04}.
It was found that the effect gives only a minor contribution to the
force and for this reason it could be estimated separately for a
homogeneous material neglecting the grain boundaries.

In the local limit for the optical conductivity the expression similar to
Eq. (\ref{cond}) was found \cite{Sot03}:

\begin{equation}
\sigma \left( \omega \right) =\sigma _{D}\left( \omega \right) \left(
1-3\alpha \left[ \frac{1}{2}-\alpha +\alpha ^{2}\ln \left( 1+1/\alpha
\right) \right] \right) ,  \label{Optcond}
\end{equation}

\noindent where the conductivity in the Drude model $\sigma _{D}\left(
\omega \right) $ and the function $\alpha =\alpha \left( \omega \right) $
are defined as

\begin{equation}
\sigma _{D}\left( \omega \right) =i\frac{\varepsilon _{0}\omega
_{p}^{2}}{%
\omega +i\omega _{\tau }},\quad \alpha \left( \omega \right)
=i\frac{v_{F}/D%
}{\omega +i\omega _{\tau }}\frac{{\cal R}}{1-{\cal R}}.  \label{alph}
\end{equation}

\noindent The longitudinal component of the electric field, which is
perpendicular to the film surface, does not feel the grain boundaries
because
the grains are columnar. Therefore, Eq. (\ref{Optcond}) is true only for
the transverse
field component. For this reason one has to distinguish two dielectric
functions in the model: the longitudinal function $\varepsilon _{l}\left(
\omega \right) =\varepsilon \left( \omega \right) $, which coincides with
the dielectric function of single crystalline gold, and the transverse
function $\varepsilon _{t}\left( \omega \right) $. The latter one can be
presented in the following form \cite{Sot03}:

\begin{equation}
\varepsilon _{t}\left( \omega \right) =\varepsilon \left( \omega \right) +3
\frac{\omega _{p}^{2}}{\omega \left( \omega +i\omega _{\tau }\right)
}\alpha \left[ \frac{1}{2}-\alpha +\alpha ^{2}\ln \left( 1+1/\alpha \right)
\right] .
\label{epst}
\end{equation}

Using these dielectric functions Sotelo et al. \cite{Sot03}
predicted the film reflectivity that was compared with the
measured reflectivity of the films with different mean grain
size $D$. The experimental data were
fitted in the visible and near infrared range to find the only parameter $%
{\cal R}$. This parameter was estimated as ${\cal R}=0.650\pm
0.025$. The results for different grain sizes all fall within the indicated
errors. Prediction of the model was compared with the reflectivity in the
infrared and excellent agreement was found.

\subsection{The Casimir force as a function of the voids fraction and the
grain size\label{Sec5.2}}

To estimate influence of the film preparation procedure on the
Casimir force, we use two parameter model taking into
account the main reasons for the deviation of the optical
properties from the bulk material. Voids in the film
are described with the effective medium approximation (see Eq.
(\ref{EMA})) where the volume fraction of voids, $f_{V}$, is
considered as a free parameter. The other source of significant
variation of the optical properties is the electron scattering
on the grain boundaries. As was shown above this effect is completely defined
by the mean grain size $D$ if the reflection coefficient of electron,
${\cal R}$, is fixed. The grain size $D $ is the second free parameter
which can be easy measured with AFM for any given film. Because the
contribution of each effect, voids or grains, to the Casimir force
is estimated on the level of a few percents, they can be considered
separately. To make the calculations we have to fix the dielectric
function of single crystalline gold. Instead of linear interpolation
between $\varepsilon ^{\prime \prime }\left( \omega _{1}\right) $
and $\varepsilon ^{\prime \prime }\left( \omega _{0}\right) $ in
the Drude tail region, as was done for the upper limit evaluation,
more realistic quadratic interpolation on the linear plot is used
to fit the minimum $\varepsilon ^{\prime \prime }\left( \omega
\right) $ that happens at $\omega \approx 2\ eV$ for all the data.
The real part of dielectric function, $\varepsilon ^{\prime }\left(
\omega \right) $, does not show any special features in the Drude
tail region. That is why for
$\varepsilon ^{\prime }\left(\omega \right) $ the Drude function is
used below $\omega_{1}$ and the handbook data \cite{The70,HB1}
above $\omega _{1} $.
Defined in this way function $\varepsilon ^{\prime }\left( \omega
\right) $ shows much smaller discontinuity at $\omega =\omega
_{1}$ than in the handbook between Dold and Mecke data
\cite{Dol65} and Th\`{e}ye data \cite{The70}. It should be
stressed that these details in the dielectric function behavior
can influence the force only on the level of 0.1\%.

\begin{figure}[tbp]
\includegraphics[width=8.6cm]{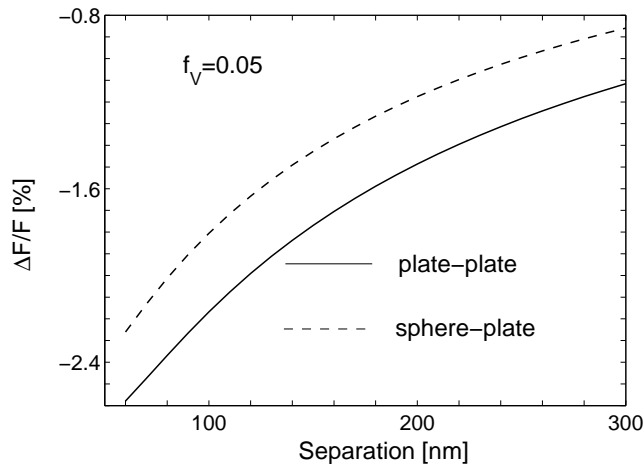}\newline
\caption{The relative correction to the force due to presence of voids in
the film as a function of separation. The results are given for the void
fraction $f_V=0.05$. The dependence of the correction on $f_V$ is
approximately linear.} \label{fig8}
\end{figure}

It is relatively easy to calculate the change in the force due to variation
of the void fraction in the film. It is assumed that the void fraction is
small and one can use Eq. (\ref{SolEMA}) for the effective dielectric
function $\left\langle \varepsilon \left( \omega \right) \right\rangle $.
This function is substituted in the dispersion relation (\ref{K-K}) to find $
\left\langle \varepsilon \left( i\zeta \right) \right\rangle $. The
effective dielectric function at imaginary frequencies is used then in the
Lifshitz formula to find the Casimir force. The relative correction to the
force, $\Delta F/F$, as a function of distance between bodies for $
f_{V}=0.05 $ is shown in Fig. \ref{fig8} for the sphere-plate geometry
(dashed line) and plate-plate geometry (solid line). It was found that the
dependence on the void fraction can be considered as linear at least up to
$f_{V}=0.15$. Hence, the graph can be used to estimate the correction
for any reasonable value of $f_{V}$. The correction is given for the case
when the temperature effect is negligible (plasma model prescription for
the $n=0$ term). The thermal correction can be estimated using Table
\ref{tab2}.

As was discussed in Sec. \ref{Sec3.1} all unannealed samples
demonstrate presence of voids. The void fraction for the best
unannealed sample evaporated with e-beam on $NaCl$ substrate was
estimated as $f_{V}=0.04$ \cite{Asp80} assuming that annealed
Th\`{e}ye films had no voids. For Johnson and Christy films
\cite{Joh72} $f_{V}$ was estimated as 0.09 and the same value is
realized for e-beam evaporated films in Wang et al. experiment
\cite{Wan98}. We expect that $f_V=0.04$ is the smallest void fraction in
unannealed samples because this value is realized for the films growing
epitaxially \cite{Asp80}. In the AFM experiment \cite{Har00} at the
minimal separation $a=63\ nm$\ the correction to the force for the
best unannealed sample is estimated then as $-1.8\ \%$. In the
MEMS experiment \cite{Dec03b}, where the smallest separation was
$a=260\ nm$, for the correction it is found $-1.0\ \%$. These
values practically coincide with that presented in Ref.
\cite{Sve03b}, where the calculation was performed using as the
perfect material the handbook data extrapolated to low frequencies
according to Ref. \cite {Lam00}. This coincidence shows once more that
the force is much more sensitive to the variation of the Drude
parameters than to any other details of the dielectric function.
Here and in Ref. \cite{Sve03b} the same value of the plasma frequency,
$\omega _{p}\approx 9\ eV$, was used for the perfect material.

It is not so straightforward to find the correction to the force
due to variation of the grain size. The reason of
complication is that the effect is described
by two different dielectric functions $\varepsilon _{l}\left(
\omega \right) $ and $\varepsilon _{t}\left( \omega \right) $. In this case
the reflection coefficients $r_{s}$ and $r_{p}$ have to be expressed via
the surface impedances $Z_{s}$ and $Z_{p}$ \cite{Kli68}. Discussion of
the impedance approach to the Casimir force evaluation rose
controversy in the literature
\cite{Bez02,Moc02,Sve03a,Gey03,Esq03,Esq04,Sve04a}. The problem
was connected with the use of the approximate Leontovich impedance
\cite {Bez02,Sve03a,Gey03} for the force calculation. This
approach was criticized \cite{Moc02} on the basis that, in
general, one has to use two different impedances one for each
polarization and it was demonstrated that the right local
impedances reproduced the Lifshitz formula. Recently the problem
was settled finally in Ref. \cite{Esq04} where the classical
theory of nonlocal impedances \cite{Kli68} was applied to the
Casimir problem. It was demonstrated that the Leontovich impedance
is a good approximation for the propagating fields but it fails
to describe the evanescent fluctuations that give significant
contribution to the force. Calculating the correction to the force
we follow the method developed in Ref. \cite{Esq04}.

The Casimir force can be calculated using the same Lifshitz formulas (\ref
{Fpp}) and (\ref{Fsp}) where the reflection coefficients are expressed via
the impedances:

\begin{equation}
r_{s}=\frac{\zeta -Z_{s}\sqrt{\zeta ^{2}+q^{2}c^{2}}}{\zeta
+Z_{s}\sqrt{%
\zeta ^{2}+q^{2}c^{2}}},\quad r_{p}=\frac{\sqrt{\zeta
^{2}+q^{2}c^{2}}%
-Z_{p}\zeta }{\sqrt{\zeta ^{2}+q^{2}c^{2}}+Z_{p}\zeta }.
\label{rimp}
\end{equation}

\noindent In general, the impedances are functions of frequency
$\zeta $ and wave vector component along the plate $q$. They can
be written via the dielectric functions \cite{Kli68} and
analytically continued to the imaginary frequencies \cite{Esq04}:

\begin{equation}
Z_{s}\left( i\zeta ,q\right) =\frac{2}{\pi }\frac{\zeta }{c}%
\int\limits_{0}^{\infty }\frac{dk_{z}}{\left( \zeta ^{2}/c^{2}\right)
\varepsilon _{t}\left( i\zeta \right) +k^{2}},  \label{Zs}
\end{equation}

\begin{equation}
Z_{p}\left( i\zeta ,q\right) =\frac{2}{\pi }\frac{\zeta }{c}%
\int\limits_{0}^{\infty }\frac{dk_{z}}{k^{2}}\left[ \frac{q^{2}}{\left(
\zeta ^{2}/c^{2}\right) \varepsilon _{l}\left( i\zeta \right) }+\frac{%
k_{z}^{2}}{\left( \zeta ^{2}/c^{2}\right) \varepsilon _{t}\left( i\zeta
\right) +k^{2}}\right] ,  \label{Zp}
\end{equation}

\noindent where $k=\sqrt{k_{z}^{2}+q^{2}}$ is the absolute value of
the wave vector. Because in our case the dielectric functions are local
(does not depend on $k$), the integrals here can be calculated explicitly.
It gives

\begin{equation}
Z_{s}=\frac{1}{\sqrt{\varepsilon _{t}+\left( cq/\zeta \right) ^{2}}},
\label{Zsex}
\end{equation}

\begin{equation}
Z_{p}=\left( \frac{cq}{\zeta }\right) \left( \frac{1}{\varepsilon _{l}}-%
\frac{1}{\varepsilon _{t}}\right) +\frac{\sqrt{\varepsilon _{t}+\left(
cq/\zeta \right) ^{2}}}{\varepsilon _{t}}.  \label{Zpex}
\end{equation}

\noindent The dielectric functions that enter $Z_{s} $ and $Z_{p} $ can
be written at imaginary frequencies with the help of Eq. (\ref{epst}) as

\[
\varepsilon _{l}\left( i\zeta \right) =\varepsilon \left( i\zeta \right) ,
\]

\begin{equation}
\varepsilon _{t}\left( i\zeta \right) =\varepsilon \left( i\zeta \right) -3%
\frac{\omega _{p}^{2}}{\zeta \left( \zeta +\omega _{\tau }\right) }\alpha
\left[ \frac{1}{2}-\alpha +\alpha ^{2}\ln \left( 1+1/\alpha \right) \right] ,
\label{epsim}
\end{equation}

\noindent where $\varepsilon \left( i\zeta \right) $ is the dielectric
function of perfect material and

\begin{equation}
\alpha \left( i\zeta \right) =\frac{v_{F}/D}{\zeta +\omega _{\tau
}}\cdot\frac{{\cal R}}{1-{\cal R}}.  \label{alphaim}
\end{equation}

\begin{figure}[tbp]
\includegraphics[width=8.6cm]{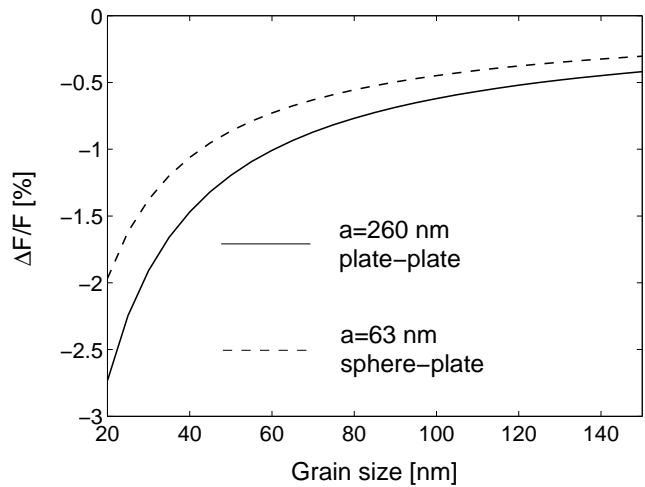}\newline
\caption{The relative correction to the force due to electron scattering on the grain
boundaries as a function of the mean grain size. The results are given for
the smallest separations in the experiments \cite{Har00} (dashed line) and
\cite{Dec03b} (solid line).} \label{fig9}
\end{figure}

To calculate the dependence of the Casimir force on the grain
size, the impedances (\ref{Zsex}) and (\ref{Zpex}) were substituted
in the reflection coefficients (\ref{rimp}) which were used then in
the Lifshitz formulas (\ref {Fpp}), (\ref{Fsp}). Correction to the force as
a function of the grain size is shown
in Fig. \ref{fig9}. The results are given for the closest separations
$a=63\ nm$ in the AFM experiment \cite{Har00} (dashed line) and $a=260\ nm$
in the last MEMS experiment \cite{Dec03b} (solid line). The correction found
here is somewhat larger than in Ref. \cite{Sve03b}. For the grain size $%
D=50\ nm$ it is estimated as -0.9\% for the AFM and -1.2\% for the
MEMS experiments in contrast with the corresponding values -0.5\%
and -0.9\% reported previously \cite{Sve03b}. The difference is
explained by the larger value of the relaxation frequency for the
perfect material, $\omega _{\tau }=0.035\ eV$, used in Ref. \cite{Sve03b}.

It was already stressed that, in general, the effect of finite grain size is
not reduced to the effective relaxation frequency which follows from Eqs. (\ref
{cond}), (\ref{cond1})

\begin{equation}
\omega _{\tau }^{eff}=\omega _{\tau }\left[ 1-\frac{3}{2}\alpha
_{0}+3\alpha
_{0}^{2}-3\alpha _{0}^{3}\ln \left( 1+\alpha _{0}^{-1}\right) \right] ^{-
1}.
\label{Omteff}
\end{equation}

\noindent However, the calculation showed that the correction to the
force with a good precision is given by the relation

\begin{equation}
\frac{\Delta F}{F}\approx A\left( a\right) \frac{\omega _{\tau
}^{eff}(D)}{
\omega _{\tau }}.  \label{Adef}
\end{equation}

\begin{figure}[tbp]
\includegraphics[width=8.6cm]{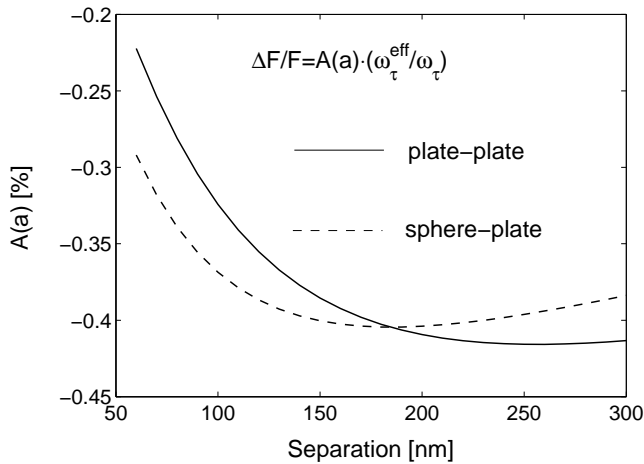}\newline
\caption{The correction due to electron scattering on the grain
boundaries as a function of separation. The function $A(a)$ is
connected with the correction by the relation (\ref{Adef}).} \label{fig10}
\end{figure}

\noindent This relation holds true at least in the investigated region $%
20<D<200\ nm$ and $60<a<300\ nm$. The function $A(a)$ is shown in
Fig. \ref{fig10} for the sphere-plate geometry (dashed line) and for the
plate-plate geometry (solid line). Therefore, the concept of the effective
relaxation frequency is happened to be good in practice.

The grain size is difficult to predict because it depends on many
details of the deposition procedure. Definitely one can say only
that for evaporated films the grain size increases with the film
thickness $h$ as $h^{0.25}$ \cite{Vaz96,Oli99}. In general, it is
not true that the grain size is close to the film thickness. For
example, the grain size for thermally evaporated film \cite{Oli99}
with the thickness of 205 nm can be estimated from the presented
image as $D\approx 50\ nm$. The image of 120 nm thick evaporated
film \cite{Liu97} gives $D\approx 40\ nm$.

The discussion above shows that instead of the physical parameters
defining the film optical properties such as the volume fraction
of voids $f_{V}$ and the grain size $D$ one can use the
effective Drude parameters connected with $f_{V}$ and $D$ by
Eqs. (\ref{Ompeff}) and (\ref{Omteff}). These effective parameters
can be extracted directly from the optical data in the infrared
as was described in Sec. \ref{Sec3.3}. The Drude
parameters are the most important information about the film that
one has to know to make high precision prediction of the Casimir
force. The details of the film dielectric response in the
near infrared and visible range can change the force only on the
level of 0.5\% or smaller.

Voids and grain boundaries are the main but not the only factors
affecting the electron system in the metallic film. The effective
mass of electron, $m_{e}^{\ast} $, can be modified by the strains
which exist in the film and change slightly the crystallographic
parameters. It would modify the Fermi surface but the changes are
not sufficient to account for the observed variation of $\omega_{p}$
 \cite{The70}. Some minor contribution, however, is not excluded.
Internal electron-scattering defects like impurities or local
defects can affect the relaxation time. This mechanism becomes
dominant over the grain boundaries in alloys. For monoatomic
material like $Au $ one can hardly believe that the local defects
(0-dimensional) can play more significant role than the grain
boundaries (2-dimensional). The Drude tail is very sensitive to the
details of the film preparation \cite{The70,Asp80}. The mechanisms
of this dependence are not well understood \cite{The70} but we have
seen that the Drude tail can change the force less that 0.5\% at the
smallest separation.

\section{Discussion\label{Sec6}}

A number of arguments were provided in this paper to show that
the optical properties of deposited gold films differ from the
properties of bulk
material on the level which cannot be ignored in the precise
evaluation of the Casimir force. The best properties demonstrate
carefully prepared annealed films but all the experiments were
made with unannealed films. The main physical reasons for the
deviation from the bulk properties are the voids in the films
reducing the density of conduction electrons and the electron scattering on
the grain boundaries significantly increasing the effective
relaxation frequency. It was demonstrated also that the Casimir
force is much more sensitive to the material optical properties in
the infrared and far infrared range than to the dielectric
response in the near infrared and visible range. For this reason
the precise values of the effective Drude parameters
are the most important information that has to be extracted from
the optical measurements for successful prediction of the force.

A reliable way to get this information is the direct optical
characterization of the films used for the Casimir force
measurement. Ordinary ellipsometry in the visible range,
$1.5-4.5\ eV$, can be helpful to see the quality of the film but
cannot be used to get the Drude parameters. Both real and
imaginary parts of the dielectric function in the wavelength range
$2-30\ \mu m$ can be routinely measured with the infrared
spectroscopic ellipsometers (for recent publications using these
apparatus see \cite{Bru04,Sch04,Dar03}). The data can be fitted
with the Drude dielectric function to find the effective plasma
frequency $\omega _{p}$ and the relaxation frequency $\omega
_{\tau }$ characterizing the film. Special attention has to be
paid to the sample preparation procedure to reduce the Drude tail
and roughness of the film. Up to date the best known method for
metal deposition is the filtered arc deposition process. The gold
films deposited with this method demonstrated the bulk-like
behavior in the near infrared \cite{Ben99} without annealing. The
method is superior to evaporation or sputtering due to significant
fraction, 70-80\%, of $Au$ ions with the energy $40-50\ eV$ in the
plasma. For high precision experiments the sample preparation and
characterization is as important as the measurement of the force
itself.

While there is no direct optical data for the films used in the Casimir
force experiments, one can establish the upper limit on the force using the
dielectric function of single crystalline gold. This upper limit was
discussed in Sec. \ref{Sec4}. It was already noticed \cite{Sve00b} that
the AFM experiment \cite{Har00} gives at smallest separations the force
which is larger than the upper limit. Much more detailed analysis of the
dielectric function of perfect single crystalline gold carried out in this
paper supports this conclusion.

Let us be more precise. For illustration we
consider only one experimental point of the closest approach in the AFM
experiment, but the same conclusions will be true for all the points with
$a<67 \ nm$. The first point presented in Fig. 5 of Ref. \cite{Har00}
corresponds $a=63\ nm$ and $F_{exp }=492\ pN$. This and the other
points of close approach were found from the original postscript figure
presented in the preprint quant-ph/0005088 with the digitizing errors (0.5
pixel) $0.15\ nm$ in the separation and $0.4\ pN$ in the force. One can
compare $F_{exp }$ with the upper limit on the force which was
discussed in Sec. \ref{Sec4}. For the reduction factor it was found $\eta
\left( 63\,nm\right)=0.445$ that is equivalent to the force $F\left(
63\,nm\right) =464\ pN$.
The difference $F\left( 63\,nm\right) -F_{\exp }=-28\ pN$ should be
compared with the experimental error that was estimated as $\Delta
F_{\exp }=3.5\ pN$ \cite{Har00}.

Due to error in the absolute separation $\Delta a=1\ nm$ \cite{Har00} the
first experimental point could correspond to $a=62\ nm$. In this case the
upper limit is $\eta \left( 62\,nm\right) =0.442$ and $F\left( 62\,nm\right)
=483\ pN$. It should be stressed that the theoretical uncertainty in the
force because of $\Delta a$ is $\left[ F\left( 62\,nm\right) -F\left(
63\,nm\right) \right] /F\left( 63\,nm\right) =4.1\%$. It is considerably
larger than the claimed 1\% agreement between the theory and experiment
\cite{Har00}. This problem has been already emphasized before
\cite{Ian03}. Even for $a=62\ nm$ the upper limit is still  $9 \ pN$
smaller than the measured force.

In reality the theoretical force should be smaller than the upper
limit due to deviation of the film optical properties from the
properties of bulk material. The smallest volume fraction of
voids was estimates in Sec. \ref {Sec5.2} as 4\%. It reduces the
absolute value of the force at $a=63\ nm$ on 1.8\%. The electron
scattering on the grain boundaries reduces the force additionally.
As the largest possible value of the mean grain size for the
unannealed $87\ nm$-thick gold film we take $D=100\ nm$. According to
Eq. (\ref{cond}) it corresponds to the resistivity $4.4\ \mu \Omega \cdot
cm$ or equivalently to the effective relaxation frequency
$\omega _{\tau }^{eff}=0.044\ eV$. The minimal reduction of the force
due to finite grain size is estimated then as 0.5\%.
Therefore, in the very best situation the theoretical force is expected to be
smaller than $F\left( 63\,nm\right) \left( 1-0.018-0.005\right) =453\ pN$
or $472\ pN$ for $a=62\ nm$. The difference with the experimental value
is striking.

Recently the result of the AFM experiment \cite{Har00} has been
reanalyzed \cite{Che04}. Three of 30 scans were rejected due to
excessive noise while the rest 27 scans were used to calculate the
force and the standard deviation. Smaller value was found for the
standard deviation $\Delta F_{exp }=2.8\ pN$. The only point for
the force was reported: at $a=62\ nm$ the mean force was $F_{exp
}=486\ pN$. Both position and the force are smaller than that in
Ref. \cite {Har00} but the
authors did not make any comment in this respect. The problem with
the uncertainty of absolute separation $\Delta a$ \cite{Ian03} was
dismissed on the basis that the experiment is in a good agreement
with the theory. The theoretical result for the force was found as
in Ref. \cite{Lam00} using the handbook data and the Drude
parameters $\omega _{p}=9.0\ eV$, $\omega _{\tau }=0.035\ eV$ to
extrapolate to the low frequencies. We have seen in Sec.
\ref{Sec4} that this procedure gives the force very close to the
upper limit. The films used in the experiments should have
different Drude parameters. In this respect the authors
\cite{Che04} stated that $\omega _{p}$ is determined by the
properties of the elementary cell which cannot be influenced by
the sample quality. This statement is wrong because, as shows
Eq. (\ref{Omp}), $\omega _{p}$ depends on the concentration of the
conduction electrons which, indeed, depends on the sample quality. The
voids will reduce the electron concentration in the sample.
Even more confusing is the
discussion of the sample dependence of $\omega _{\tau }$. The
authors introduce not imaginary but real characteristic frequency
$\omega _{ch}=c/2a$ and assume that this frequency gives the main
contribution to the force. As was discussed in detail in Sec. \ref{Sec2}
low frequencies $\omega\ll \omega_{ch}$ give the main contribution to
the dispersion integral (\ref{K-K}) for $\varepsilon \left( i\zeta \right)$.
This fact was already stressed in a number of publications
\cite{Bos00b,Sve00b,Ian03,Ian04}. Because of this mistake the
authors \cite{Che04} took into account some minor effects in the
near infrared like electron-electron scattering that gave
quadratic frequency dependence of the relaxation frequency
\cite{The70} but missed very important and well established effect
like significant increase of the film resistivity in comparison with the
bulk material.

The same analysis cannot be done for the latest MEMS experiment
\cite{Dec03b} mainly because of huge roughness correction. In
contrast with the AFM experiment, where the roughness correction was
smaller than 0.5\%, in this experiment the roughness correction is
as large as 25\%. One can hardly believe that the calculations based
on the local separation \cite{Dec03b} can be true for the shape
dependent Casimir force with the precision on the level of  1\%. In
the experiment \cite{Dec03b} one of the plates was covered with
copper but in the present paper an extensive analysis of gold was
made. For this reason one can say only that the sample effect should
reduce the force on the level of  2\%. It is much larger than the
declared experimental errors. The roughness correction and sample
effects do not allow us to support the conclusion \cite{Dec03b} on
the possibility to distinguish between different approaches to the
temperature correction.

\section{Conclusions\label{Sec7}}

In this paper we analyzed the sample dependence of the Casimir force.
It was already noted \cite{Lam99,Sve00a} that the optical
properties of metallic films used to measure the force can deviate
significantly from the bulk material but here, for the first time,
serious analysis of the effect based on the facts about gold films was
undertaken. It was stressed that the behavior of the dielectric
function of a metal in the far infrared, where direct optical data are
inaccessible, is crucial for the precise evaluation of the force.
The Casimir force depends on the dielectric function $\varepsilon
\left( i\zeta \right) $ connected with the observable function
$\varepsilon ^{\prime \prime }\left( \omega \right) $ via the
dispersion relation (\ref {K-K}). Because $\varepsilon ^{\prime
\prime }\left( \omega \right) $ is large at low frequencies, the
far infrared range gives the most significant contribution in
$\varepsilon \left( i\zeta \right) $ and, therefore, in the force.
For example, at $\zeta =1\ eV$ the contribution of the real
frequency domain $\omega <0.125\ eV$ to $\varepsilon \left( i\zeta
\right) $ was estimated as 75\%.

Sample dependence of the optical properties of gold films was
discussed in Sec. \ref{Sec3}. The data for 40 years of
investigations have been collected and analyzed. These data clearly
showed that significant difference between samples cannot be
explained by the
experimental errors and had to be attributed to genuine sample
dependence of the optical properties. The mechanisms of this
dependence have been clarified in the literature. Variation of
$\varepsilon ^{\prime \prime }\left( \omega \right) $ in the
visible range was connected with the volume fraction of voids
$f_{V}$ in the films. The void fraction is especially significant
in unannealed films. For the best unannealed films which start to grow
epitaxially on $NaCl$ substrate $f_{V}$ was estimated as 4\% but
in the worst cases it could be more than 20\%. The fraction of
voids is very important for the Casimir force because it defines
the precise value of the plasma frequency. The concentration of
conduction electrons in the sample decreases with the increase of
$f_{V}$ and, as the result, $\omega _{p}$ decreases.

Analysis of the data available in the infrared range revealed the
same tendency of the dependence on the sample preparation
procedure. It was demonstrated that at $\omega <1\ eV$ the data
for real and imaginary parts of $\varepsilon \left( \omega \right)
$ could be reasonably well described with the Drude model. The
fitting procedure allowed one to find the effective Drude parameters
for the investigated samples (see Table \ref{tab1}). Surprisingly,
the handbook data \cite{HB1} in the range $\omega <1\ eV$ taken
from the old paper by Dold and Mecke \cite{Dol65} happen to be
bad. The best fit gave $\omega _{p}=7.52\ eV$, $\omega _{\tau
}=0.061\ eV$. It means that the films were poorly prepared and
were full of voids and defects for the electron scattering. On the
other hand bulk $Au(110)$ and well annealed gold films
demonstrated much better characteristics close to that typically
used for the calculation of the Casimir force \cite{Lam00}: $\omega
_{p}=9.0\ eV$, $\omega _{\tau }=0.035\ eV$.

It was stressed that the force was measured between unannealed
films for which one can hardly expect that the effective Drude
parameters coincide with that for the bulk material. However,
using the "ideal" dielectric function of gold one can establish
the upper limit on the Casimir force. This dielectric function was
carefully described in Sec. \ref{Sec4}. The upper limit happened
to be very close to the force evaluated in Ref. \cite {Lam00}.
The use of nearly the same $\omega_{p} $ and $\omega_{\tau} $ for
"ideal" gold in this paper and in Ref. \cite{Lam00} explains the
coincidence of the forces. Thus, the result of Ref. \cite{Lam00}
should be considered rather as the upper limit on the force than
as the force between real films.

The expected deviations of the force from the case of perfect single
crystalline gold were analyzed in Sec. \ref{Sec5}. An appreciable
correction to the Casimir force was found due to the presence of voids
in the film. Even for the smallest fraction of voids, $f_{V}=0.04$, the
relative correction to the absolute value of the force in the AFM
experiment \cite{Har00} was found to be -1.8\% at the smallest
separation $a=63\ nm$. For the MEMS experiment \cite{Dec03b} at
$a=260\ nm$ the correction was -1.0\%. The other effect that was
estimated was influence of the grain boundaries in the film on
the electron scattering.
This effect is the main reason for the increase of the film
resistivity or the effective relaxation frequency. For the mean
grain size of 50 nm (typical for 100 nm thick film) the
corrections due to the finite grain size were estimated as -0.9\%
for the AFM experiment and -1.2\% for the MEMS experiment at the
smallest separations in each case. Both corrections cannot be
ignored in the experiment where the force was measured with the
precision on the level or better than 1\%. A number of additional
factors can affect the dielectric function. Not all of them are
well understood but one can indicate strains in the film or
some internal electron-scattering defects. All additional effects are
expected to reduce the absolute value of the force less than 1\%.

The best way to get the precise values of the Drude parameters
is the direct measurement of the dielectric functions of the film
used in a Casimir force experiment. The most important frequency
range where this function should be known is the infrared range
$\omega <1\ eV$. It can be done with the infrared spectroscopic
ellipsometers. Special attention has to be payed to the film
preparation procedure.

\end{document}